\documentclass[a4paper, 11pt]{article}

\usepackage[pdftex]{graphicx}
\usepackage[cmex10]{amsmath}
\usepackage{amssymb}
\usepackage{booktabs}
\usepackage{url}
\usepackage{bbm}
\usepackage[margin=2cm]{geometry}

\newcommand{\bK}{\mathbf{K}}
\newcommand{\bY}{\mathbf{Y}}
\newcommand{\by}{\mathbf{y}}
\newcommand{\bX}{\mathbf{X}}
\newcommand{\bx}{\mathbf{x}}
\newcommand{\bbeta}{\boldsymbol \beta}

\newcommand{\beps}{\boldsymbol \varepsilon}
\newcommand{\real}{\mathbb{R}}
\newcommand{\mN}{\mathcal{N}}

\begin{document}

\title{Probabilistic Forecasting of Temporal Trajectories of Regional Power Production---Part 1: Wind}
\author{Thordis~L.~Thorarinsdottir,
Anders~L{\o}land, and
Alex~Lenkoski\thanks{Norwegian Computing Center, Oslo, Norway (e-mail: thordis@nr.no).}
}

\maketitle

\begin{abstract}
Renewable energy sources provide a constantly increasing contribution
to the total energy production worldwide. However, the power
generation from these sources is highly variable due to their
dependence on meteorological conditions.  Accurate forecasts for the
production at various temporal and spatial scales are thus needed for
an efficiently operating electricity market. In this article -- part 1 
-- we
propose fully probabilistic prediction models for spatially aggregated
wind  power production at an hourly time
scale with lead times up to several days using weather forecasts from
numerical weather prediction systems as covariates.  After an
appropriate cubic transformation of the power production, we build up a
multivariate Gaussian prediction model under a Bayesian inference
framework which incorporates the temporal error correlation. In an application to predict wind 
production in Germany, the method provides calibrated and skillful
forecasts.  Comparison is made between several formulations of the
correlation structure.  
\end{abstract}

\section{Introduction}

Recent years have seen a worldwide proliferation in
energy production from renewable energy sources.  In Germany, for
instance, renewable energy accounted for $36.0\%$ of the total national
energy production in 2017 compared to $6.6\%$ in 2000 according to the
Arbeitsgemeinschaft Energiebilanzen, a working group founded by energy
related associations in Germany. This increase is to a large extent
due to expansion in wind and photovoltaic (PV) solar power
production. (Wind power production accounted for $1.6\%$ in 2000 and as much as
$17.6\%$ in 2017.)
However, as these energy sources rely on the prevailing wind and solar
irradiance conditions, as well as other weather variables, the
resulting power generation is highly variable and uncertain.
Simultaneously, accurate production forecasts are needed for the
management of electricity grids, for scheduling of the production at
conventional power plants as well as for general decision making on
the energy market e.g.\ \cite{LangeFocken2006,elberg&2015}.  These different
contexts imply varying loss functions which, together with the need to
control the trade-off between risk and return, calls for a
probabilistic forecasting framework  \cite{Gneiting2011}.  
Probabilistic forecasts are becoming increasingly frequent for wind
power forecasting \cite{Bremnes2004, PinsonKariniotakis2010, JeonTaylor2012, Pinson2012, Pinson2013,Hong&2016,Dowell&2016}.  

Time series approaches e.g.\ \cite{Gneiting&2006, Pinson2013,Dowell&2016} usually outperform other
methods for lead times up to 3-6 h after which they may be improved
upon by statistical methods that relate the expected production to
weather forecasts from numerical weather prediction (NWP) models.  The
usual approach is to model a single unit or a farm.
\cite{Taylor&2009} utilize the local wind speed observations to
calibrate wind speed density forecasts which are subsequently
transformed to wind power while \cite{Bremnes2004} and
\cite{PinsonMadsen2009} directly model the relationship between the
wind speed forecasts and the power production. Alternatively,
\cite{Messner&2013} employ an inverse power curve transformation in a
regression framework and \cite{JeonTaylor2012} consider a stochastic
power curve model.  A recent comprehensive review of available wind
power prediction models at various time scales is
\cite{GIEBEL201759}.  The Global Energy Forecasting Competitions
(GEFCom2012 \cite{Hong&2014} and GEFCom2014 \cite{Hong&2016}) have
 attracted hundreds of participants worldwide, who contributed many novel ideas
to the energy forecasting field, and day ahead wind power forecasting
in particular. A clear majority of the contestants applied machine learning
methods, like gradient boosting regression  and quantile regression
forest \cite{nagy&2016} or K-nearest neighbors \cite{zhang2016}. 

Many end-users require forecasts of aggregated power production over a
market region or for a regional transmission organization. Regional
forecasts are often formed by an upscaling of a set of individual
sites \cite{LangeFocken2006, SiebertKariniotakis2006}. This requires an up-to-date account of the
overall installed capacity, hourly production data for the region as a
whole and production data from a representative set of sites. In
countries such as Germany with continued expansion of renewable energy
production, this can be a cumbersome task. Instead, we propose to
directly predict the aggregate country-wide production using spatially
averaged NWP forecasts of the relevant weather variables as inputs.
Similarly, applications in system operation and planning call for
forecasts over multiple lead times returning calibrated forecast
trajectories. Several studies have applied copula approaches to
account for the error correlation structure across lead times
\cite{Pinson&2009} or the correlation between different locations
\cite{Hagspiel&2012, Louie2014, PapaefthymiouKurowicka2009}. However,
the marginal predictive distributions are often modeled independently
in a non-parametric fashion e.g.\ \cite{Pinson&2009}.  We specify
the probabilistic prediction model as a
Bayesian hierarchical model, which allows us to incorporate a
correlation structure in both the model parameters associated with
each lead time  as well as the error structure across lead
times. A recent review \cite{Engeland&2017} notes that renewable energy forecasting systems
that focus on long lead-times, regional level data and use the combination of meteorological and production data is
largely unexplored in the literature, making this system one of the first to combine these aspects.

The NWP forecasts and the German power production data are introduced
in the next Section~\ref{sec:data}.  The prediction models and the
statistical inference methods are derived in
Section~\ref{sec:models}.  The forecast verification methods we employ
are
described in Section \ref{sec:verification}, and the results are
presented in Section~\ref{sec:results}. We conclude with a discussion in
Section~\ref{sec:discussion}.

\section{Forecast and observation data}\label{sec:data}

We employ the NWP forecast ensemble issued by the European Centre for Medium-Range Weather Forecasts (ECMWF) which has been shown to perform well in this setting \cite{giebel2011state}.   The 50-member ECMWF ensemble system operates at a global horizontal resolution of $0.25 \times 0.25$ degrees, a resolution of approximately $32 \times 32$ km over Germany, and a temporal resolution of 3--6 h with lead times up to ten days \cite{LeutbecherPalmer2008, Molteni&1996}. We restrict attention to the forecast initialized at 00:00 UTC, corresponding to 2:00 am local time in summer and 1:00 am local time in winter, and lead times up to 72 h for 100 m wind speed.  

The hourly wind power production data for Germany are obtained from the European Energy Exchange (EEX) where they are available to all members that trade on the EEX, see \url{www.transparency.eex.com/en/}.  We use data from the calendar year 2011 to assess the optimal length of the training period for parameter estimation as well as for determining the prior parameters of the Bayesian model.  Given these values, we then test our methods on data from 2012.  In order to obtain equally long training periods for all dates, data from the previous year is used for the parameter estimation at the beginning of a year. 

The differences between the individual members in an ECMWF ensemble stem from random perturbations in initial conditions and stochastic physics parameterizations in the numerical model.  The ensemble members are thus statistically indistinguishable, or exchangeable, and should be given equal weights in a regression framework.  We therefore reduce the ensemble to a single forecast given by the ensemble average.  For the operation and management of electricity grids, power production predictions are needed on an hourly basis.  However, for the first 72 h, the ECMWF forecasts have a temporal resolution of 3 h.  We derive hourly forecasts through a spline interpolation conditional on the variables being non-negative.  

In a third preprocessing step, we aggregate the forecasts in space by taking the spatial average. The wind power production is largely concentrated in the northern half of the country.  Rather than employing the aggregated forecasts over the entire country, we thus focus on the northern half only (latitudes greater than 51) for the wind speed, which results in a stronger relationship between the forecasts and the power production. As a result, the wind speed forecast is an average over 371 grid locations. 

\section{Wind power prediction model}\label{sec:models}

\begin{figure*}[ht]
\centering
\includegraphics[width=1\textwidth]{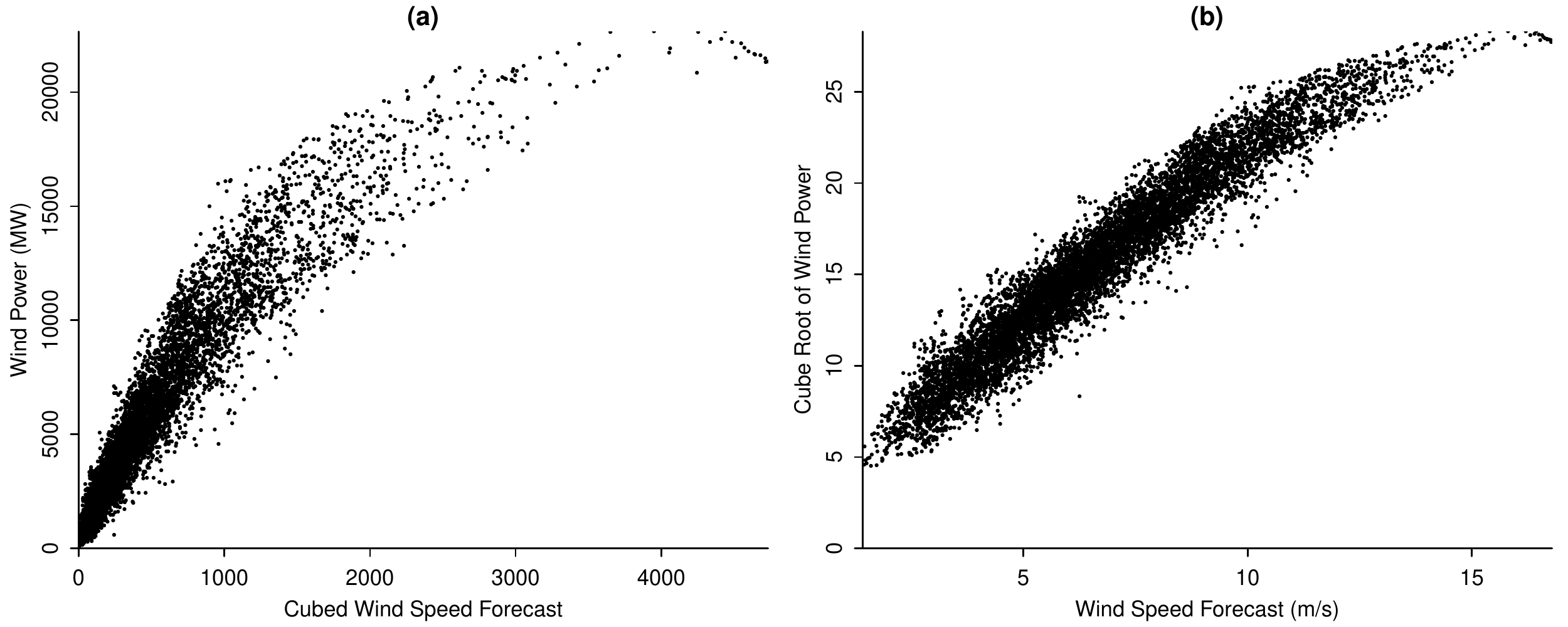}
\caption{Relationship between hourly wind power production in Germany in 2011 and the corresponding wind speed forecasts: (a) wind power production against 1-24 h cubed wind speed forecasts, and (b) cube root of the wind power production against 1-24 h wind speed forecasts.}\label{fig:wind cubed}
\end{figure*}

The nonlinear relationship between wind speed and the power output from an individual turbine is described by the power curve. The turbine blades begin to rotate at the cut-in speed and the maximum power output of the turbine is generated from the rated speed until the cut-out speed, at which speed the blades stop rotating to prevent damage.  These parameters may vary between different turbines and, in practice, the power curve is not deterministic \cite{JeonTaylor2012}.  Our production data is the aggregated power output from thousands of wind turbines spread over a large geographic area. It is thus highly unlikely that the wind speed is below the cut-in speed or above the cut-out speed simultaneously at all the turbines. This is confirmed by Figure~\ref{fig:wind cubed} which shows that the data does not appear heavily censored.  

For wind speed values between the cut-in speed and the rated speed, the power output from an individual turbine is generally proportional to the cubed wind speed \cite{HeringGenton2010}.  As shown in Figure~\ref{fig:wind cubed}(a), the relationship between the cubed average wind speed forecasts and the aggregated wind power production is highly heteroskedastic with a larger spread for higher wind speeds.  We thus follow \cite{Messner&2013} and, in our prediction model, we model the relationship between the average wind speed forecasts and the cube root of the resulting wind power, see Figure~\ref{fig:wind cubed}(b).  Denote by $\bx_w = (x_{w1}, \ldots, x_{wT})^\top$ the wind speed forecast for lead times $1$ to $T$. The wind power production $Y_{wt}$ at time $t \in \{ 1, \ldots, T\}$ is then given by 
\begin{equation}\label{eq:wind model}
Y_{wt}^{1/3} = \beta_{0t} + \beta_{1t} x _{wt} + \beta_{2t} x_{wt}^3 + \varepsilon_{t},  
\end{equation}   
where $\beta_{it} \in \real$ for $i = 0, 1,2$ and the error terms $\beps = (\varepsilon_1, \ldots, \varepsilon_T)^\top$ are assumed correlated in time,
\begin{equation}\label{eq:wind error}
\beps \sim \mN_T (0, \bK^{-1}),
\end{equation}
for some precision matrix $\bK = \{K_{ij} \}_{i,j =1}^T$. The particular form of the regression equation in \eqref{eq:wind model} was selected based on the average marginal predictive performance for 1-24 h forecasts in 2011 (results not shown). Alternatives included regression equations with one to three covariates from the set $\{ \bx_w, \bx_w^2, \bx_w^3\}$.  The power production is inherently nonnegative and the normal assumption in \eqref{eq:wind error} might therefore be physically unrealistic. However, as the predictand in \eqref{eq:wind model} never takes values close to zero, see Figure~\ref{fig:wind cubed}(b), we find that, in practice, the predicted probability of negative production is negligible.  Further model validation criteria are discussed in Section~\ref{sec:results} below. 

Denote by $\bX = [ \mathbb{I}_T~\textup{Diag}(\bx_w)~\textup{Diag}(\bx_w^3)]$ the $T \times 3T$ joint covariate matrix for $Y^{1/3}_{w1},\ldots,Y^{1/3}_{wT}$ based on the model in \eqref{eq:wind model}. Here, $\mathbb{I}_T$ is the identity matrix of size $T$ and $\textup{Diag}(\bx)$ denotes a diagonal matrix with $\bx$ on the diagonal.  The likelihood model for $\bY = (Y_{w1}^{1/3},\ldots,Y_{wT}^{1/3})^\top$ is then given by 
\begin{equation}\label{eq:wind likelihood}
\bY \sim \mathcal{N}_T(\bX \bbeta, \, \bK^{-1}), 
\end{equation}
where $\bbeta = (\bbeta_0^\top, \bbeta_1^\top, \bbeta_2^\top)^\top$ with $\bbeta_{i} = (\beta_{i1}, \ldots, \beta_{iT})^\top$ for $i = 0,1,2$. We estimate the parameters under a Bayesian inference framework with conjugate prior distributions of the form
\begin{align}\label{eq:priors}
\bbeta \, | \, \bK_0, \mathbf n_0 & \sim \mathcal{N}_{3T} \Big(\boldsymbol 0,  \big[ \textup{Diag}(\mathbf n_0) \otimes \bK_0 \big]^{-1} \Big), \\
\bK & \sim \mathcal{W}_G (3, \mathbb{I}_T), \label{eq:prior K} \\
\bK_0 & \sim \mathcal{W}_{G_0} (3, \mathbb{I}_T), \label{eq:prior K0} \\
n_{0i} & \sim \Gamma (1, 0.5), \quad i = 0, 1, 2,
\end{align}
where $\otimes$ denotes the Kronecker product, $\mathbf n_0 = (n_{00}, n_{01}, n_{02})^{\top} \in \real_+^3$ and the gamma distribution is parameterized in terms of shape and rate.  The three vectors $\bbeta_0$, $\bbeta_1$ and $\bbeta_2$ are thus assumed independent under the prior and the inflation factors $\mathbf n_0$ account for the potential variation in the scale of the covariates.  

The conjugate prior distribution for the precision matrix $\bK$  is the G-Wishart distribution $\mathcal{W}_G$ \cite{Roverato2002}, where we use the notation of \cite{Lenkoski2013}. The support of $\mathcal{W}_G$ is the space of all symmetric positive definite matrices which fulfill the conditional independence structure given by the graph $G = (V,E)$ where $V = \{1, \ldots, T\}$ and $E \subset V \times V$.  That is, $K_{ij} = 0$ whenever $(i,j) \notin E$. For instance, if $G$ is the conditional independence structure of an autoregressive process of order $1$, AR(1), then it holds that $(i,j) \in E$ if and only if $|i -j| \leq 1$.  The autoregressive structure is, however, completely flexible and may vary over time with the prior parameters in \eqref{eq:prior K} providing a slight shrinkage towards no autocorrelation to prevent potential overfitting.  If $G$ is the independence graph with $(i,j) \in E$ if and only if $i=j$, the prior distribution in \eqref{eq:prior K} is equivalent to a $\Gamma(3/2, 1/2)$ prior distribution on each marginal precision. 

\subsection{Full model}\label{sec:full model}

Under the full model, we simultaneously estimate the marginal predictive distributions and the error correlation.  Here, we set $\bK_0 = \bK$, implying a weakly informative prior, with $G = G_0$ the conditional independence structure of an AR(1) process.  Let us assume that $N$ forecast-observation pairs are available. In order to obtain samples from the joint posterior distribution of $\bbeta$ and $\bK$ given the data, we iteratively sample from the full conditional distributions 
\begin{align}\label{eq:full conditionals}
\bbeta \, & | \, \bK, \mathbf n_0, \{\by_{n}\}_{n=1}^N, \{\bX_{n}\}_{n=1}^N  \sim \mathcal{N}_{3T}\big( \tilde{\bbeta}, \tilde{\bK}^{-1}\big), \\ 
\bK \, & | \, \bbeta, \mathbf n_0, \{\by_{n}\}_{n=1}^N, \{\bX_{n}\}_{n=1}^N \sim \mathcal{W}_G (6 + N, \mathbb{I}_T + \mathbf{S}), \label{eq:full conditional K} \\
n_{0i} \, & | \, \bbeta_i, \bK  \sim \Gamma \Big( \frac{T+2}{2}, \frac{\bbeta_i^\top \bK \bbeta_i}{2} \Big), \quad i = 0,1,2, \label{eq:full conditional n0} 
\end{align}
where
\begin{align*}
\tilde{\bK} & = \big[\textup{Diag}(\mathbf n_0) \otimes \bK \big] + \sum_{n=1}^N \bX_{n}^{\top} \bK \bX_{n}, \\
\tilde{\bbeta} & = \tilde{\bK}^{-1} \sum_{n=1}^N \bX_{n}^{\top} \bK \by_{n} , \\ 
\mathbf{S} & = \sum_{n=1}^N (\by_{n} - \bX_{n} \bbeta) (\by_{n} - \bX_{n} \bbeta)^{\top} +  \sum_{i = 0}^2 n_{0i} \bbeta_i \bbeta_i^{\top}.
\end{align*}
While it is straight forward to sample from the distributions in \eqref{eq:full conditionals} and \eqref{eq:full conditional n0}, we employ the direct sampler of \cite{Lenkoski2013} to obtain samples from the G-Wishart distribution in \eqref{eq:full conditional K}. Given the posterior parameter samples and the current wind speed forecast, we then obtain samples from the posterior predictive distribution for the wind power production by sampling a value from the likelihood model in \eqref{eq:wind likelihood} for each posterior parameter sample and transforming these to wind power. 

\subsection{Two-stage copula model} 

An alternative model construction is a two-stage Gaussian copula model which builds on the work of \cite{DobraLenkoski2011} and \cite{Moeller&2013}.  In the first stage, we perform joint estimation of the marginal predictive distributions following the set up above with $G$ equal to the independence graph.  If $G_0$ is equal to the independence graph, the marginal predictive distributions are estimated independently, while an AR(1) structure in the graph $G_0$ imposes an autogressive structure on each of $\bbeta_i$ for $i = 0, 1,2$.  We consider both of these options. 

To estimate the error correlation, we proceed as follows. The estimation of the marginal predictive distributions yields forcast-observation pairs $\{ F_{tn}, y_{tn}\}$ for $n = 1,\ldots,N$ and $t = 1,\ldots,T$, where $F$ denotes the predictive distribution. We may then infer $N$ latent Gaussian observations $\{ \mathbf z_{n} \}_{n=1}^N$ by setting $z_{tn} = \Phi^{-1}(F_{tn}(y_{tn}))$, where we denote the standard Gaussian cumulative distribution function by $\Phi$. The latent Gaussian data has likelihood
\[
p(\{ \mathbf z_{n} \}_{n=1}^N \, | \, \mathbf K_Z) = (2 \pi)^{TN/2} | \mathbf K_z |^{N/2} \exp \Big( - \frac{1}{2} \textup{tr}(\mathbf K_z, \mathbf U) \Big),
\]
where $\mathbf U = \sum_{n=1}^N \mathbf z_n \mathbf z_n^\top$ and $\mathbf K_z$ is an $N \times N$ precision matrix. Under a prior distribution of the form \eqref{eq:prior K}, the posterior distribution for $\mathbf K_z$ is thus given by 
\begin{equation}\label{eq:copula correlation posterior}
\bK_z \, | \, \{\mathbf z_n \}_{n=1}^N \sim \mathcal{W}_G (3 + M, \mathbb{I}_T + \mathbf U ). 
\end{equation}

Finally, a sample $\hat{\mathbf y}$ from the posterior predictive distribution for the wind power production is obtained in three steps:
\begin{enumerate}
\item Sample $\hat{\bK}_z$ from \eqref{eq:copula correlation posterior}.
\item Sample $\mathbf z^*$ from $\mathcal{N}_T(0, \hat{\bK}_z^{-1})$ and set $\hat{z}_t = z^*_t / \sqrt{(\hat{\bK}_z^{-1})_{tt}}$ for $t = 1, \ldots, T$. 
\item Set $\hat{y}_t = F_t^{-1}(\Phi(\hat{z}_t))$ for $t = 1,\ldots,T$, where $F_t^{-1} (u) := \max \{ y \, : \, F_t(y) \leq u \}$.     
\end{enumerate}
Here, $F_t$ denotes the marginal predictive distribution at time $t$.  Note that the latent Gaussian vector in step 2 is normalized as the inverse of $\hat{\mathbf K}_z$ which may be a covariance matrix rather than a correlation matrix.  

\section{Forecast verification methods\label{sec:verification}}

We apply various forecast verification methods for probabilistic predictions with the aim of assessing which method provides the sharpest predictive distributions subject to calibration \cite{GneitingBalabdaouiRaftery2007}. A forecasting method is calibrated if events predicted to happen with probability $p \in [0,1]$ are also realized with empirical relative frequency $p$. Calibration of univariate forecasts may be assessed empirically by plotting histograms of the probability integral transform (PIT) $F(y)$ for a predictive distribution $F$ and the corresponding realized obervation $y$ over a large set of forecast cases. For a calibrated forecast, the PIT histogram will have a uniform (flat) shape \cite{Dawid1984}. Alternatively, calibration and sharpness can be assessed directly for a fixed $p$ by calculating the average coverage and width of the corresponding prediction interval.

For assessing multivariate calibration, we calculate the multivariate rank of an observed temporal trajectory $\mathbf y = (y_1, \cdots, y_T)$ in an ensemble with $\mathbf y$ and $m-1$ samples $\hat{\mathbf y}_1, \cdots, \hat{\mathbf y}_{m-1}$ from the multivariate predictive distribution $\mathbf F$. Here, we use the band depth ranking of \cite{Thorarinsdottir&2016}. That is, we first apply a pre-rank function $\rho: \real^T \rightarrow \real_+$ given by
\begin{equation}\label{eq:bdr}
\rho(\mathbf y) = \frac{1}{T} \sum_{t=1}^T \big[m - \textup{rank}(y_t) \big] \big[ \textup{rank}(y_t) - 1\big] + (m-1),  
\end{equation}
where $\textup{rank}(y_t)$ denotes the standard univariate rank of $y_t$ in $(\hat{y}_{1t},\cdots,\hat{y}_{(m-1)t},y_t)$. The multivariate rank of $\mathbf y$ is then given by the univariate rank of $\rho(\mathbf y)$ in $(\rho(\hat{\mathbf y}_1), \cdots, \rho(\hat{\mathbf y}_{m-1}), \rho(\mathbf y))$. The calibration may now be assessed empirically by plotting the histogram of the ranks of $\rho(\mathbf y)$ over multiple forecast cases with a uniform shape indicating a calibrated forecast. Note that the definition in \eqref{eq:bdr} only holds if, with probability one, no two trajectories in the ensemble are equal, see the discussion in \cite{Thorarinsdottir&2016}.   

In addition, we calculate multiple proper scores which assess various different aspects of the predictive distribution \cite{GneitingRaftery2007}. The absolute error $| \textup{median}(F) - y |$ compares the median of the univariate predictive distribution $F$ against the observation $y$, while under the squared error $(\textup{mean}(F) - y)^2$, the mean of $F$ is the optimal point forecast \cite{Gneiting2011}. These scores are then averaged over multiple forecast cases resulting in the mean absolute error (MAE) and the root mean squared error (RMSE). Similarly, we calculate the mean continuous ranked probability score (CRPS), which compares the full distribution $F$ against the empirical distribution function of the observation $y$,
\begin{equation}\label{eq:crps}
\textup{CRPS}(F,y) = \int_{-\infty}^{+ \infty} (F(z) - \mathbbm{1}\{ x \geq y\})^2 \textup{d} z,
\end{equation}
where $\mathbbm{1}$ denotes the indicator function. To estimate the integral in \eqref{eq:crps} we employ the approximation methods described in \cite{Kruger&2016} as implemented in the {\tt R} package {\tt scoringRules} \cite{Jordan&2018}. All three scores are negatively oriented such that a smaller score indicates a better predictive performance. Furthermore, the score units are equal to that of $y$ or MWs in our case. 

\section{Results\label{sec:results}}

\subsection{Length of training period} 

We assess the influence of the amount of training data on the results by comparing the average marginal predictive performance under rolling training periods of different lengths. For wind power predictions, this is performed for the full model described in Section~\ref{sec:full model} as well as independent marginals with either an independent or AR(1) structure on the regression coefficients. Aggregated results for lead times up to 24 hours and the months of January, April, July and October of 2011 indicate that the prediction models are very robust against the amount of training data. For rolling training periods of length 50 to 150 days, the performance of all methods changes by less than 3\% when measured by the CRPS. Results for the MAE and the RMSE are similar. In the following, we use a training period of 100 days for both the marginal and the multivariate models.

\subsection{Marginal predictive performance}

\begin{figure}[!hbpt]
\centering
\includegraphics[width=0.75\textwidth]{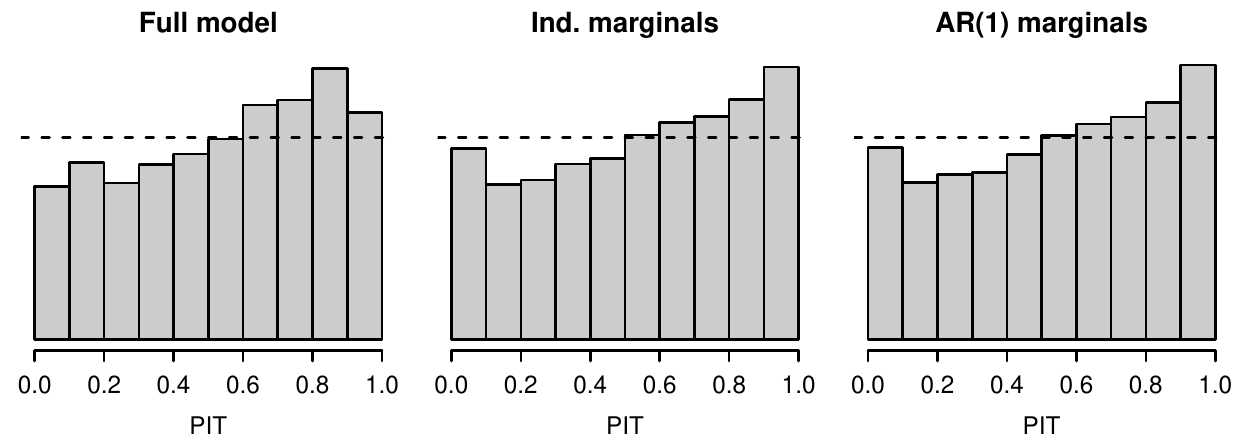}
\caption{Probility integral transform (PIT) histograms for marginal wind power predictions under three different marginal models. The PIT values are aggregated over lead times of 1-24h during the test period from January 1, 2011 to December 31, 2012, a total of 17544 forecast cases.  The dashed lines indicate the level of a perfectly flat histogram.}\label{fig:PITmarginalwind}
\end{figure}

We start by assessing the marginal predictive performance of the various models. 
The probility integral transform (PIT) historgrams (Fig. \ref{fig:PITmarginalwind}) indicate minor deviations from the ideal uniform predictive distribution, most notably a bias to the right, but not a clear under- or overdispersion. The results for the three different marginal models are very similar.

\begin{table}[!hbpt]
\centering
\caption{Calibration and sharpness of marginal predictions for wind power as measured by the coverage and width of 80\% prediction intervals.  The results are aggregated over lead times of 1-24h (Day 1), 25-48h (Day 2) and 49-72h (Day 3), and the test period from January 1, 2011 to December 31, 2012. The best results in each category are indicated in bold.
\label{tab:calib}}
\vspace{1mm}
\begin{tabular}{lcccccc}
\toprule
 & \multicolumn{3}{c}{Coverage (\%)} & \multicolumn{3}{c}{Width (MW)} \\
\cmidrule(lr){2-4} \cmidrule(lr){5-7} 
& Day 1 & Day 2 & Day 3 & Day 1 & Day 2 & Day 3 \\ 
\midrule
Full model  & {\bf 81.1} & {\bf 80.2} & {\bf 79.9} & 2964 & 3359 & 3969 \\
Ind Errors  & 76.8 & 77.3 & 78.2 & {\bf 2378} & {\bf 2932} & {\bf 3680} \\
Fully Ind  & 76.9 & 77.5 & 78.6 & 2388 & 2944 & 3692 \\
\midrule 
\end{tabular}
\end{table}

Table~\ref{tab:calib} shows the average width and coverage of 80\% prediction intervals aggregated over lead times of 1-24h (Day 1), 25-48h (Day 2) and 49-72h (Day 3). While the coverage is similar for different lead times, the width of the prediction intervals expectedly increases with lead time. The full model has somewhat wider prediction intervals and slightly better coverage than the other two models. 

\begin{table*}[ht]
\centering
\caption{Marginal predictive performance of models for wind power production as measured by mean absolute error (MAE), root mean squared error (RMSE) and mean continuous ranked probability score (CRPS). The results are aggregated over lead times of 1-24h (Day 1), 25-48h (Day 2) and 49-72h (Day 3), and the test period from January 1, 2011 to December 31, 2012. The best results in each category are indicated in bold.}\label{tab:marginal scores}
\vspace{1mm}
\begin{tabular}{lrrrrrrrrr}
\toprule
& \multicolumn{3}{c}{MAE (MW)} & \multicolumn{3}{c}{RMSE (MW)} & \multicolumn{3}{c}{CRPS (MW)} \\
\cmidrule(lr){2-4} \cmidrule(lr){5-7} \cmidrule(lr){8-10} 
& Day 1 & Day 2 & Day 3 & Day 1 & Day 2 & Day 3 & Day 1 & Day 2 & Day 3 \\
\midrule
Full model & 977 & 1117 & 1281 & 1390 & 1586 & 1801 & 676 & 778 & 902 \\ 
Ind Errors & {\bf 792} & 988 & 1185 & {\bf 1176} & 1432 & 1711 & {\bf 564} & 699 & 847 \\
Fully Ind & {\bf 792} & {\bf 987} & {\bf 1184} & {\bf 1176} & {\bf 1428} & {\bf 1707} & {\bf 564} & {\bf 698} & {\bf 846} \\
\bottomrule
\end{tabular}
\end{table*}

When the marginal predictive performance is assessed by proper scores, the marginal models perform considerably better than the full model across all scores, see  Table~\ref{tab:marginal scores}. In addition to independent and AR(1) marginals, we have also tested using higher order AR structures which yielded reduced predictive performance (results not shown). For the best model, the performance is reduced 14-24\% from day 1 to day 2 and 20-21\% from day 2 to day 3, depending on the score.

\subsection{Multivariate calibration}
\begin{figure}[!hbpt]
\centering
\includegraphics[width=0.25\textwidth]{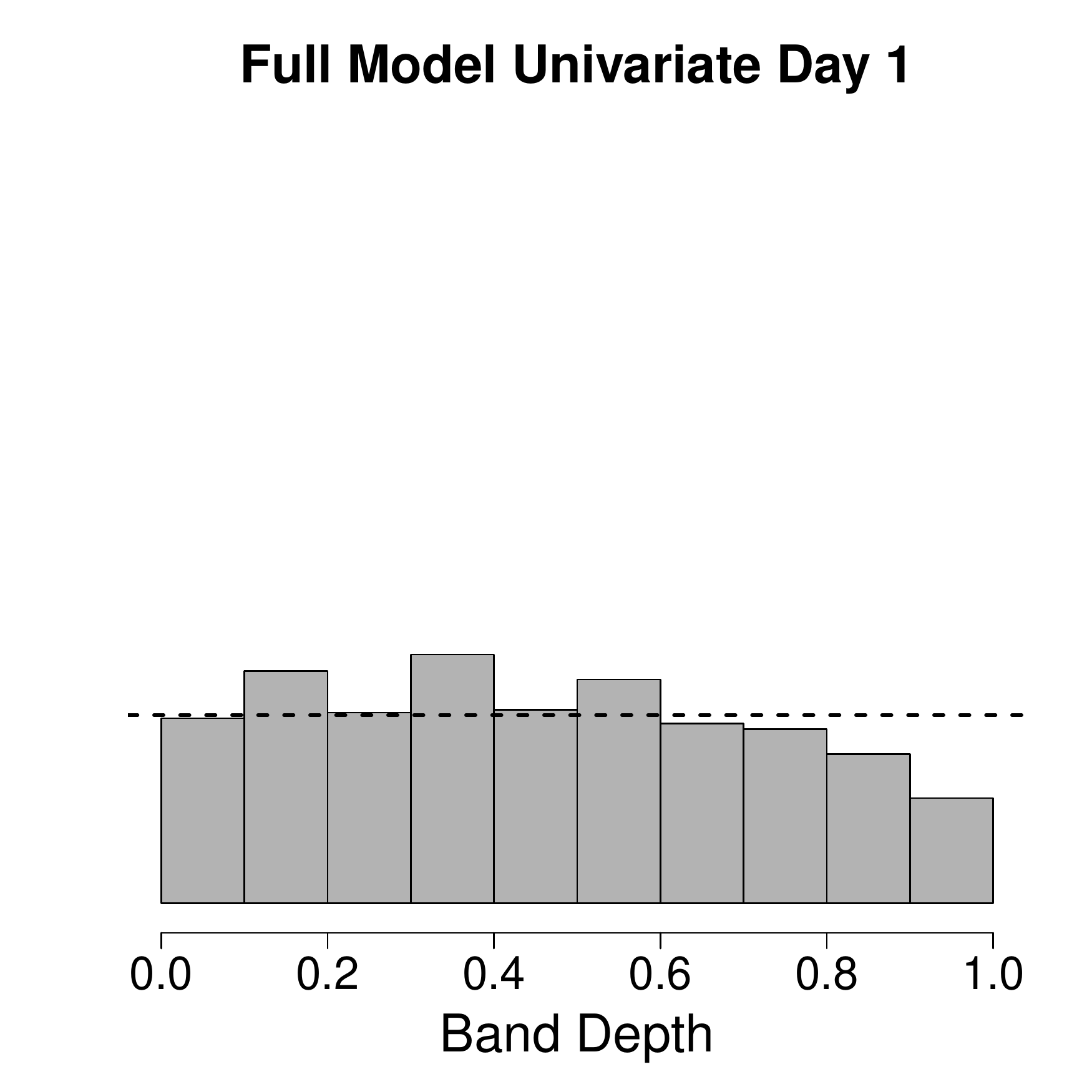}
\includegraphics[width=0.25\textwidth]{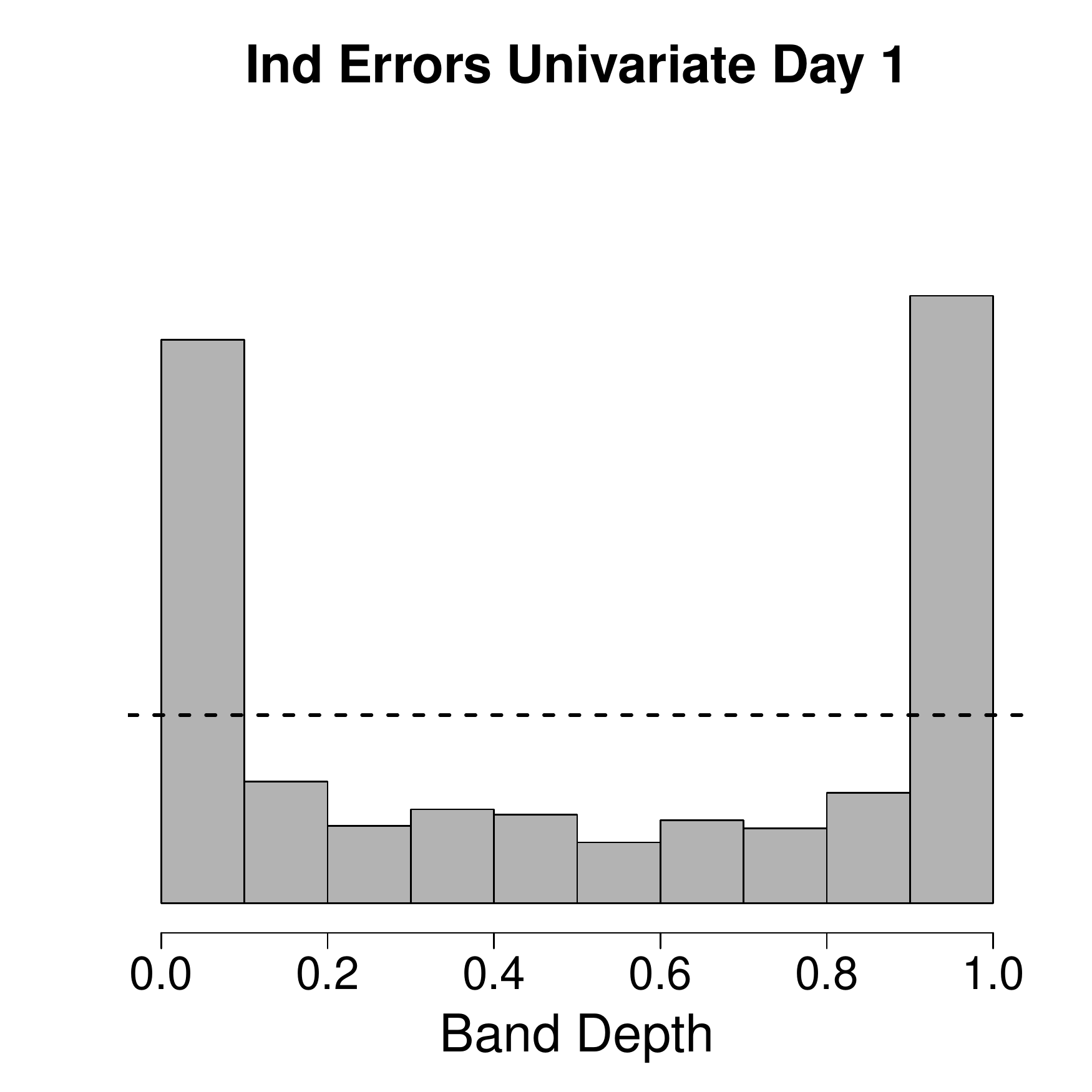}
\includegraphics[width=0.25\textwidth]{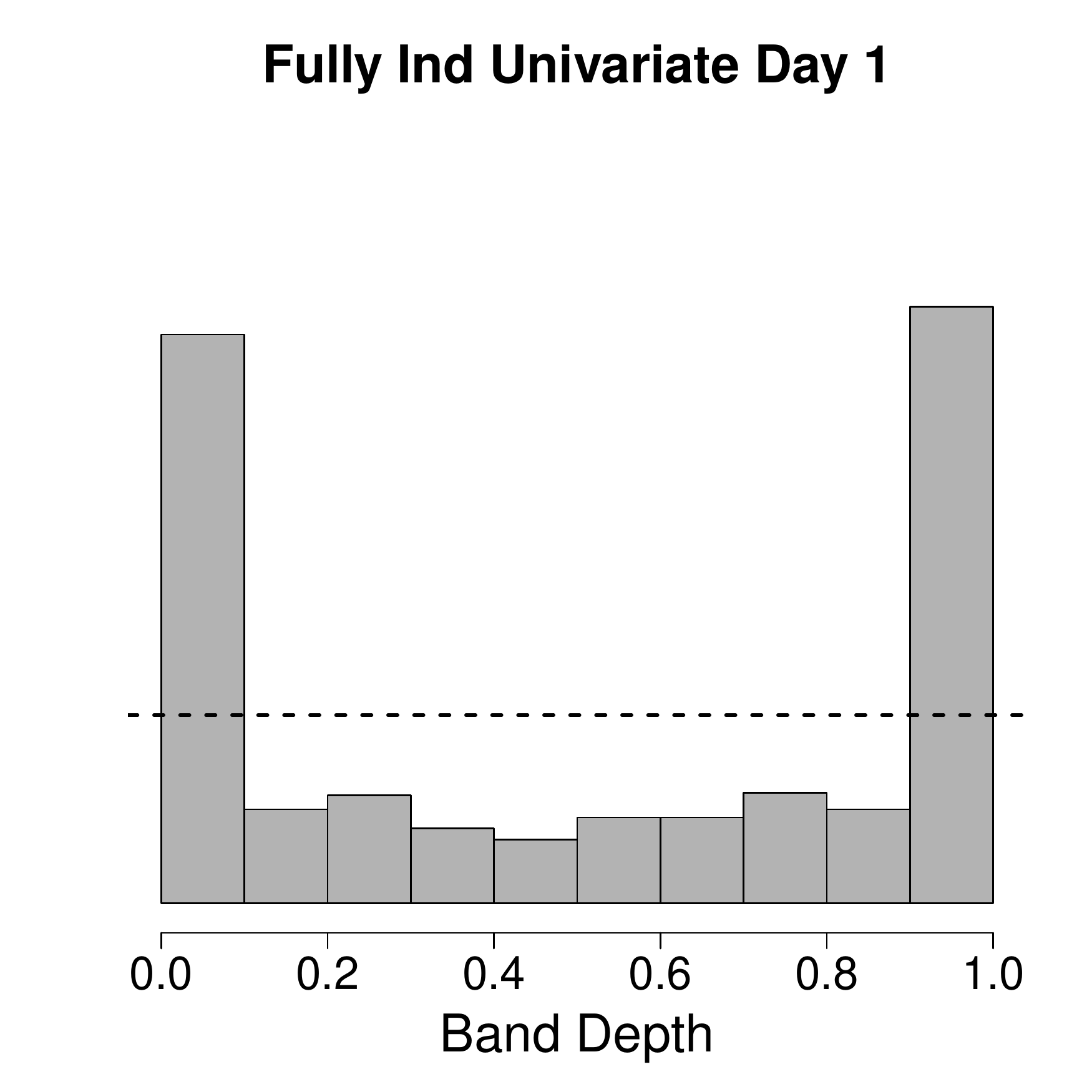}\\
\includegraphics[width=0.25\textwidth]{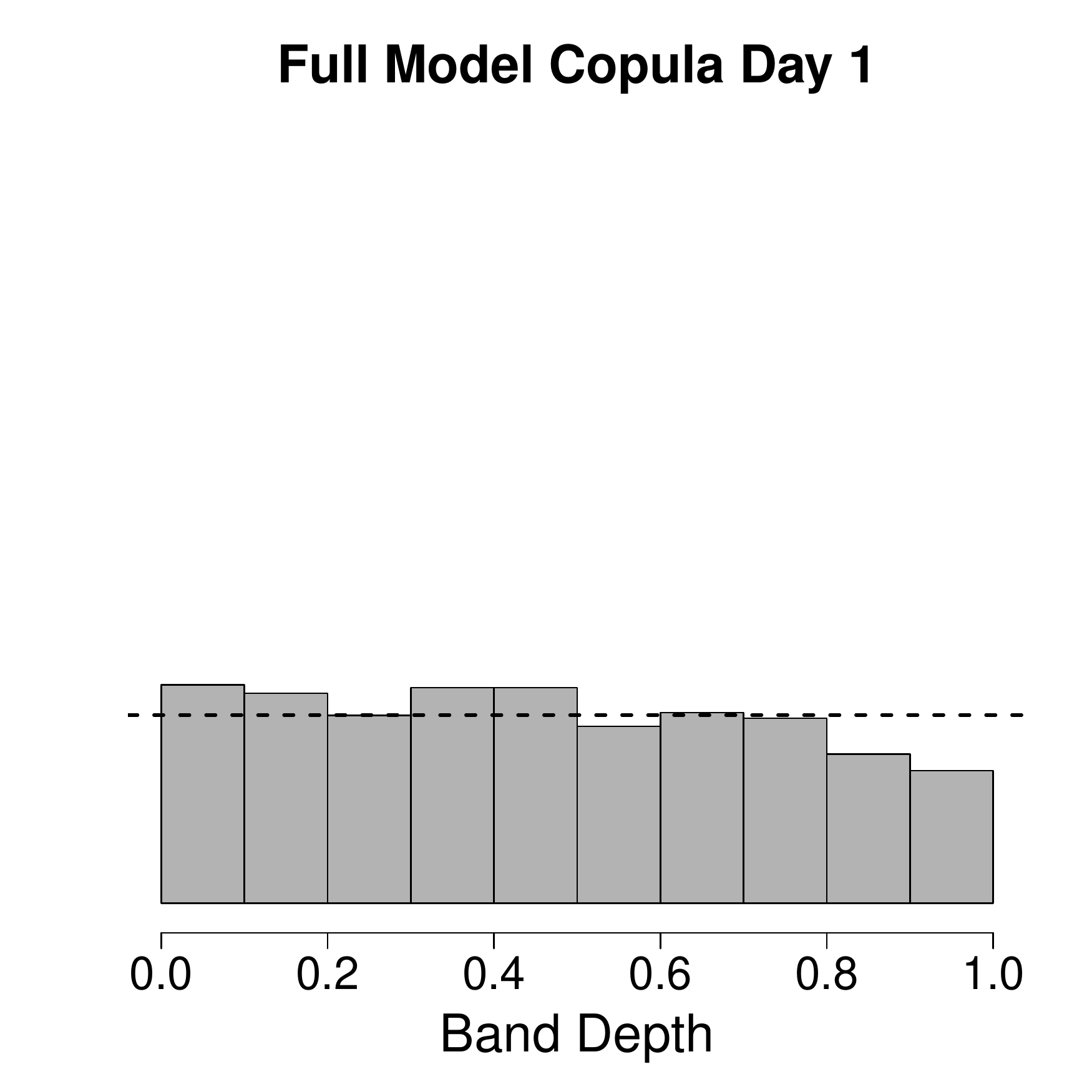}
\includegraphics[width=0.25\textwidth]{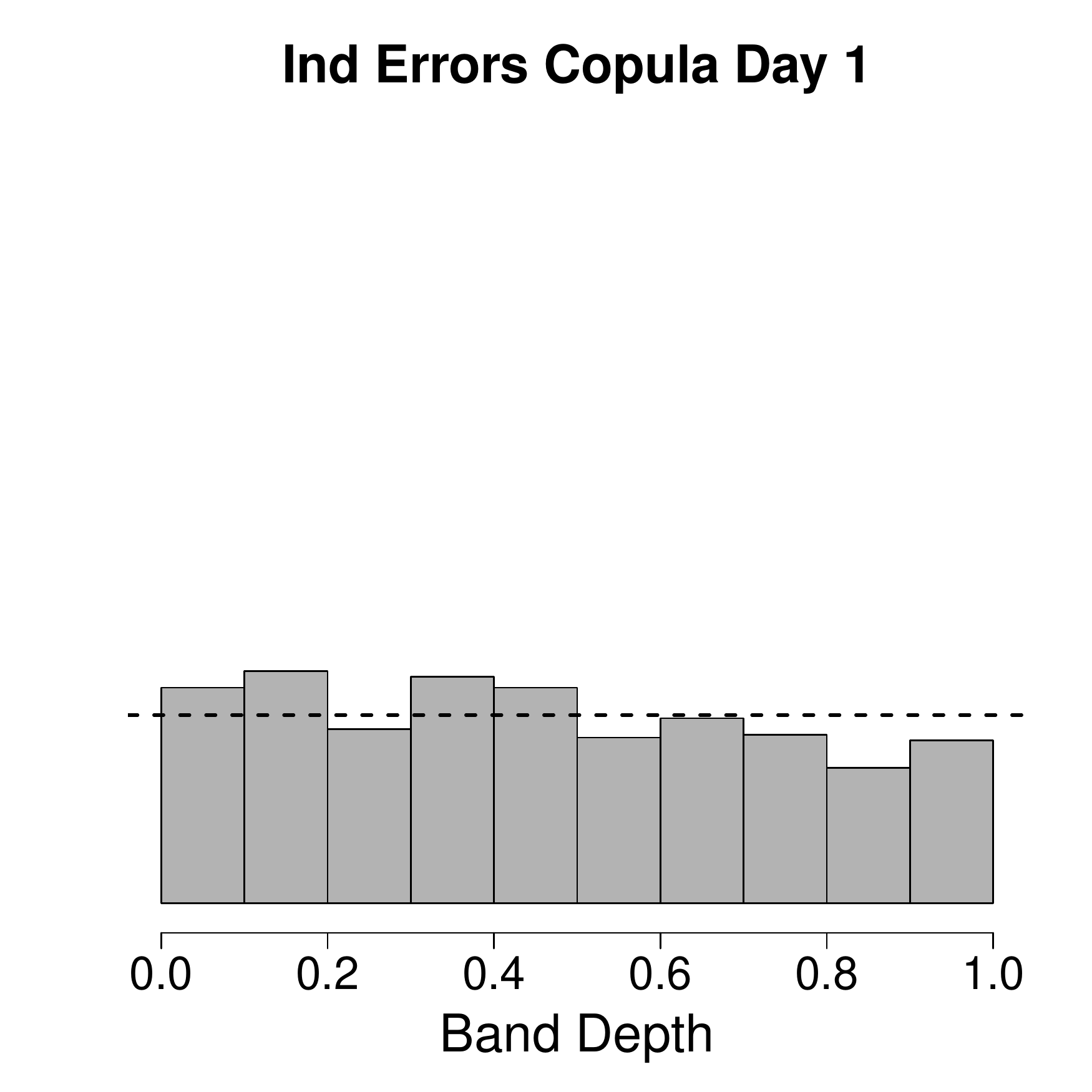}
\includegraphics[width=0.25\textwidth]{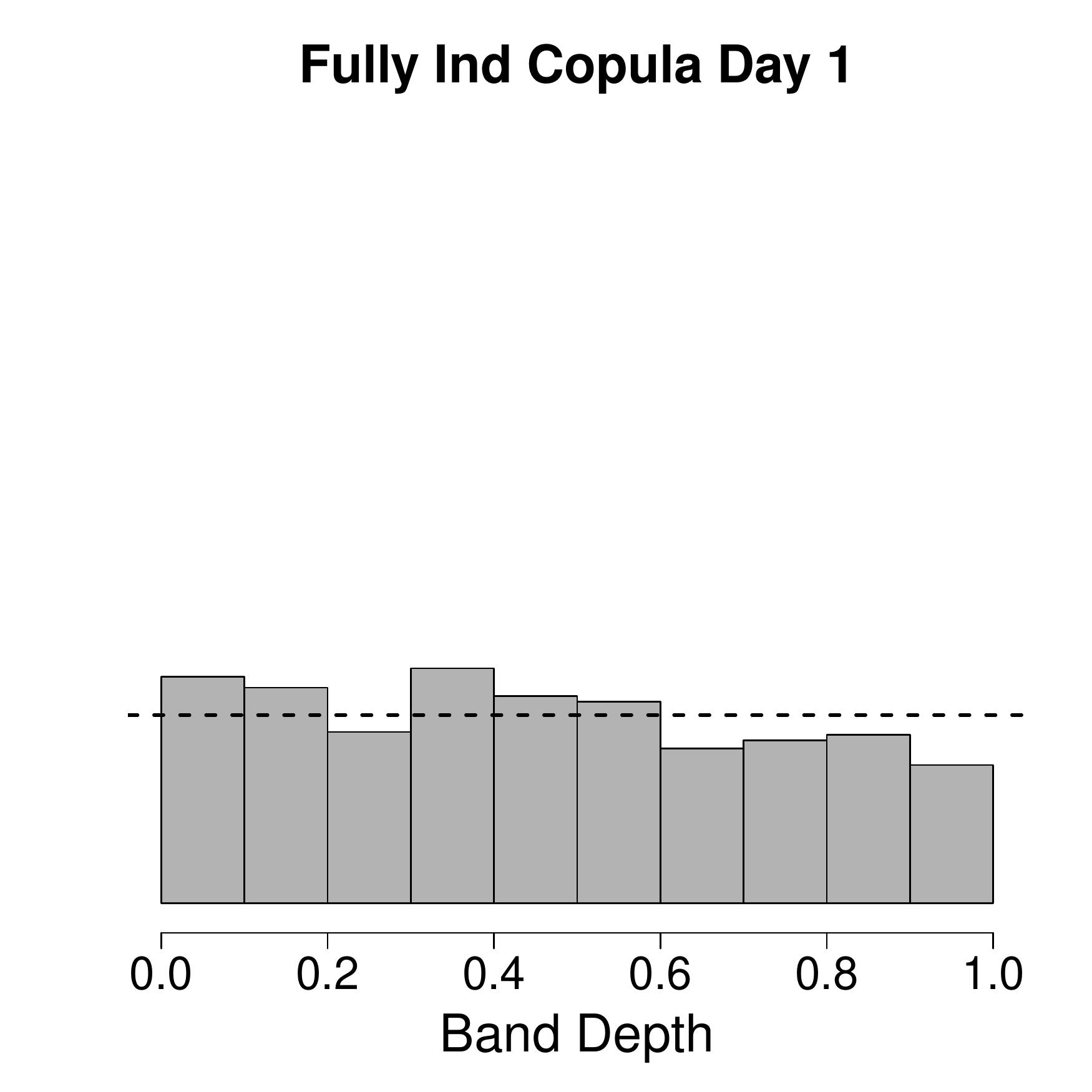}
\caption{Band depth histograms for hours 1-24 by model type (full, independent errors and fully independent respectively going from left to rigth) and under either the univariate (top row) or copula (bottom row) approach.}\label{fig:bd_wind}
\end{figure}

Figure~\ref{fig:bd_wind} shows the band depth histograms for the joint predictive distribution over hours 1-24.  For each of the three approaches, the independent model is shown on the top row while the bottom row shows the results after copula post-processing.

As expected, under the univariate approach only the full model shows evidence of near calibration, while the two approaches with completely independent errors show substantial multivariate over-dispersion.  This is unsurprising, given the strong degree that errors are correlated across hours. However, the bottom row of Figure~\ref{fig:bd_wind} shows that copula post-processing is capable of addressing these issues, and all three aproaches, after copula processing, show roughly the same degree of multivariate calibration. This suggests that it is acceptable to perform marginal inference first and then subsequently address the multivariate aspects of the forecast distribution.

\subsection{Predicting daily maxima and totals}
We now consider two distributional forecasts derived from the entire multivariate forecast. Namely, we look at the total wind speed and forecasted maximum windspeed over the 72 hours.  Since these two quantities are affected by the joint behavior of the underlying forecast, assessments of their distributional performance provides an indication of the quality of the overall joint distributional forecast.
\begin{table*}[htp]
\centering
\caption{Scores for predicting the sum of 72-hours ahead production of wind power by method}\label{tab:mult_sum}
\vspace{1mm}
\begin{tabular}{lrrrrrrrrr}
\toprule
& MAE (MW) & RMSE (MW) & CRPS (MW)\\
\midrule
Full Model Univariate & 49224 & 69786 & 34868\\
Ind Errors Univariate & \textbf{43010} & 62947 & 37712\\
Fully Ind Univariate & 43651 & 65449 & 38161\\
Full Model Copula & 49258 & 69811 & 34758\\
Ind Errors Copula & 43390 & \textbf{62851} & \textbf{30465}\\
Fully Ind Copula & 44048 & 65403 & 31066\\
\bottomrule
\end{tabular}
\end{table*}

Table~\ref{tab:mult_sum} shows the scores for each method for the sum of wind power across all 72-hours.  We see that acording to the MAE, the full model performs best, while the model with independent marginal errors followed by a copula post-processing shows the best performance.

The conclusion from Table~\ref{tab:mult_sum} is two-fold.  First, it is (self-evidently) important to have dependence in the errors in the joint distribution either by explicit direct modeling (Full model) or via subsequent copula post-processing.  Furthermore, including dependence in the regression coefficients is beneficial, which can be seen by the fact that the Fully Independent model underperforms the other methods, even after copula post processing to correlate the sampling distribution.  These results speak to the usefulness of a joint Bayesian model with dependence built into the prior.

Table~\ref{tab:mult_max} shows similar scores for the maximum.  The conclusions here are broadly in-line with those from Table~\ref{tab:mult_sum}. We see that the independent errors model with copula performs best according to MAE while the Full model with subsequent copula post processing performs best according to RMSE and CRPS.  While the ordering has changed slightly the key points hold, namely that dependence in the sampling distribution and model-imposition of dependence in regression coefficients are beneficial to predictive performance.
\begin{table*}[ht]
\centering
\caption{Scores for predicting the maximun of 72-hours ahead production of wind power by method}\label{tab:mult_max}
\vspace{1mm}
\begin{tabular}{lrrrrrrrrr}
\toprule
& MAE (MW) & RMSE (MW) & CRPS (MW)\\
\midrule
Full Model Univariate & 1340 & 1752 & 946\\
Ind Errors Univariate & 2156 & 3225 & 1679\\
Fully Ind Univariate & 2383 & 6974 & 1871\\
Full Model Copula & 1309 & \textbf{1713} & \textbf{927}\\
Ind Errors Copula & \textbf{1307} & 2051 & 935\\
Fully Ind Copula & 1499 & 6267 & 1096\\
\bottomrule
\end{tabular}
\end{table*}

\begin{figure}[!hbpt]
\centering
\includegraphics[width=0.25\textwidth]{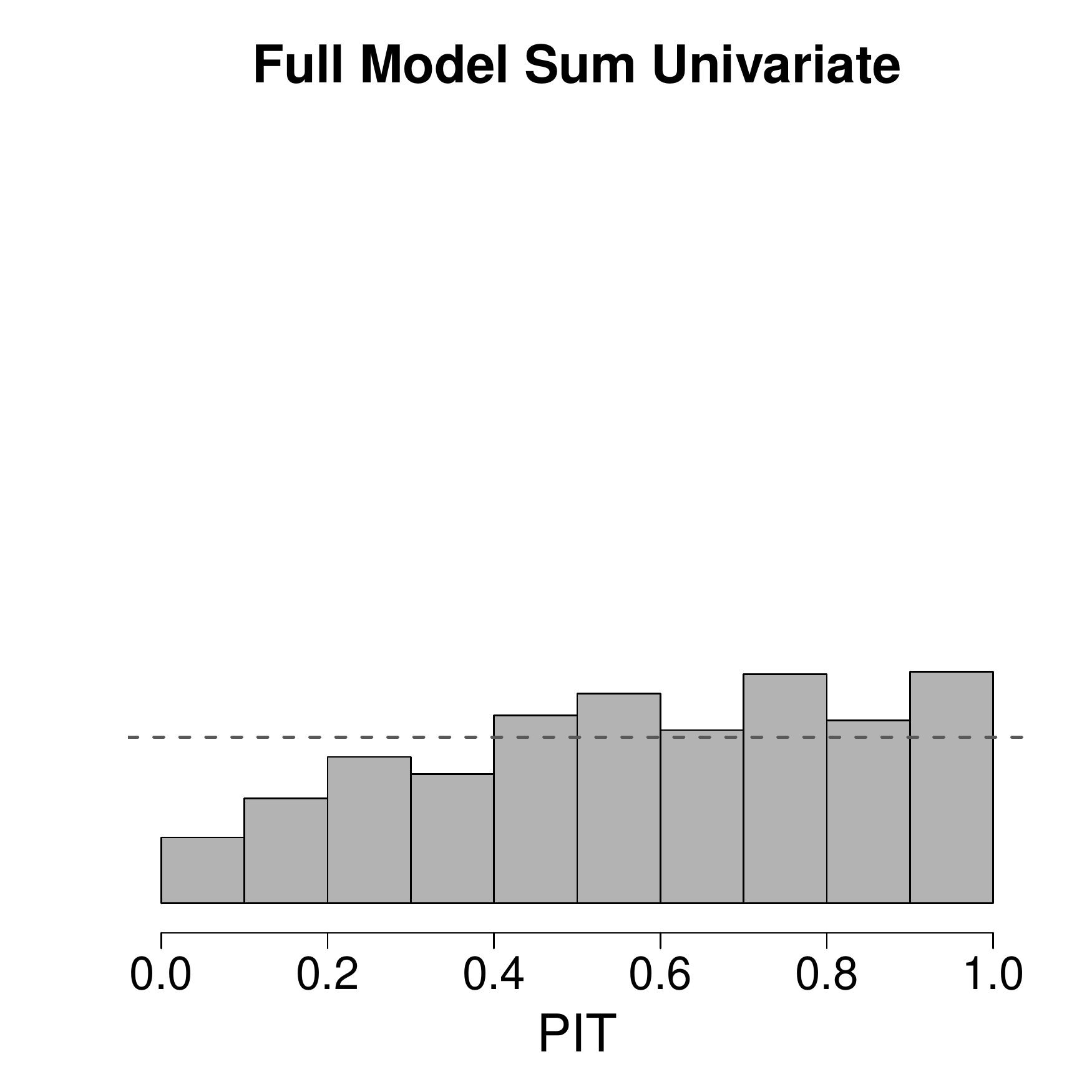}
\includegraphics[width=0.25\textwidth]{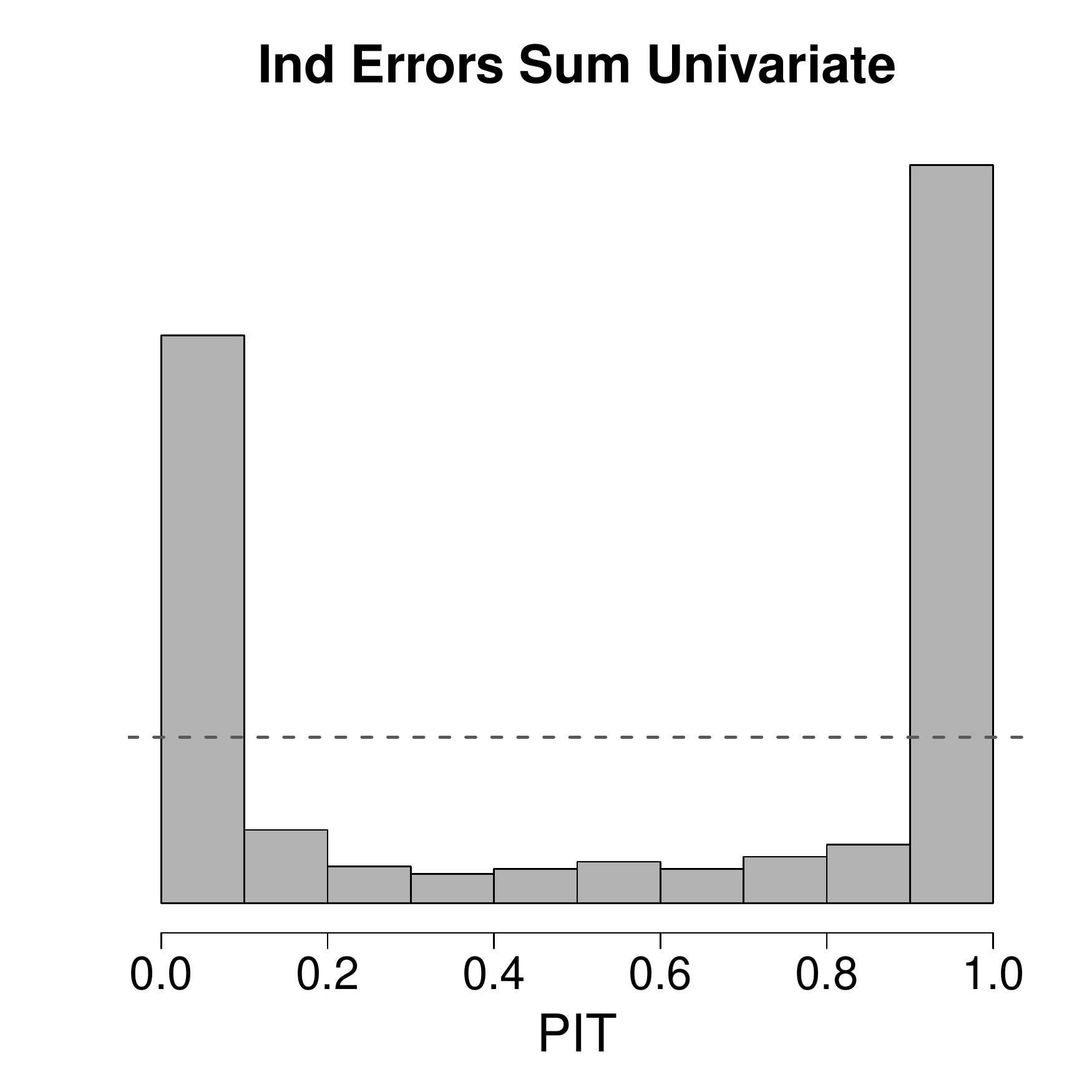}
\includegraphics[width=0.25\textwidth]{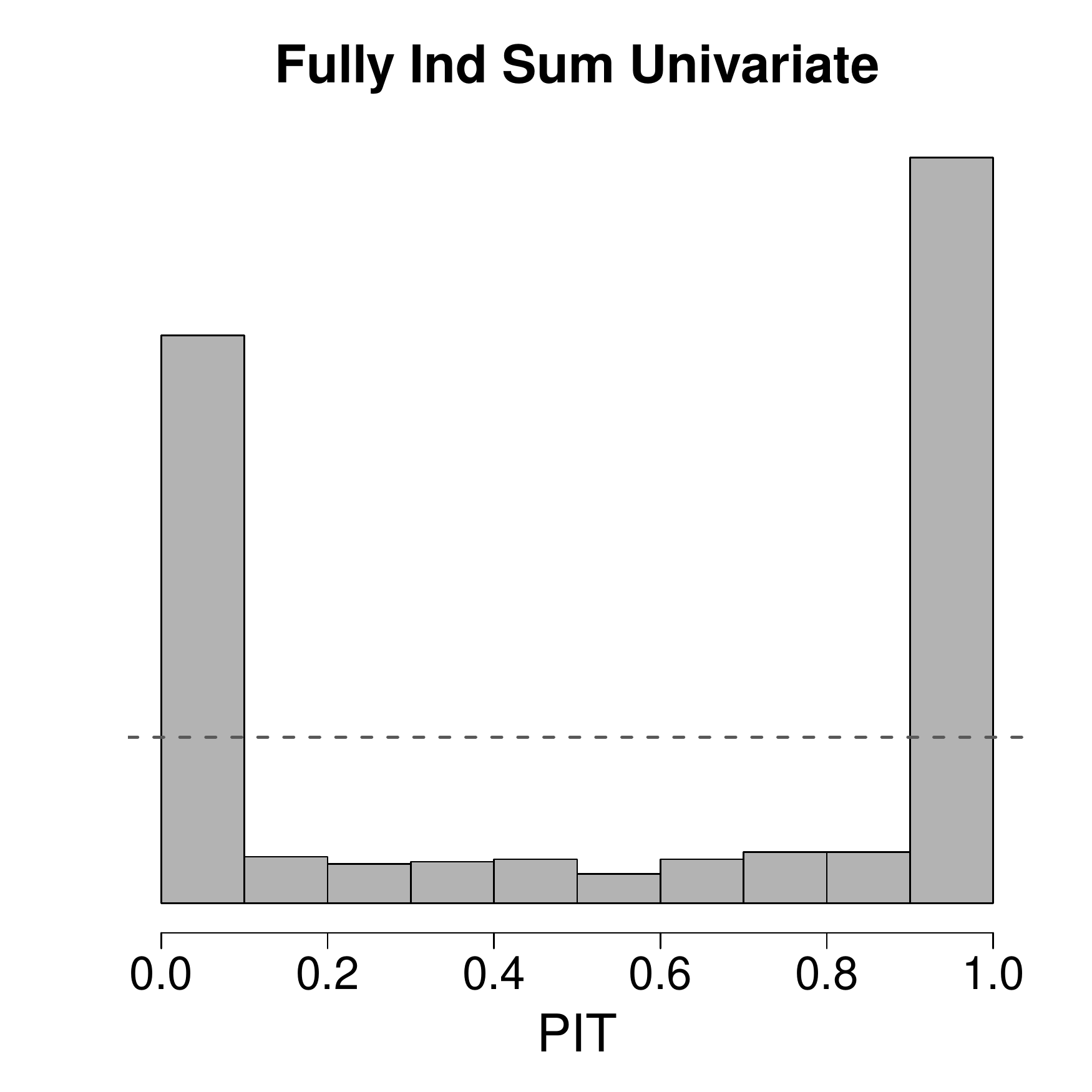}\\
\includegraphics[width=0.25\textwidth]{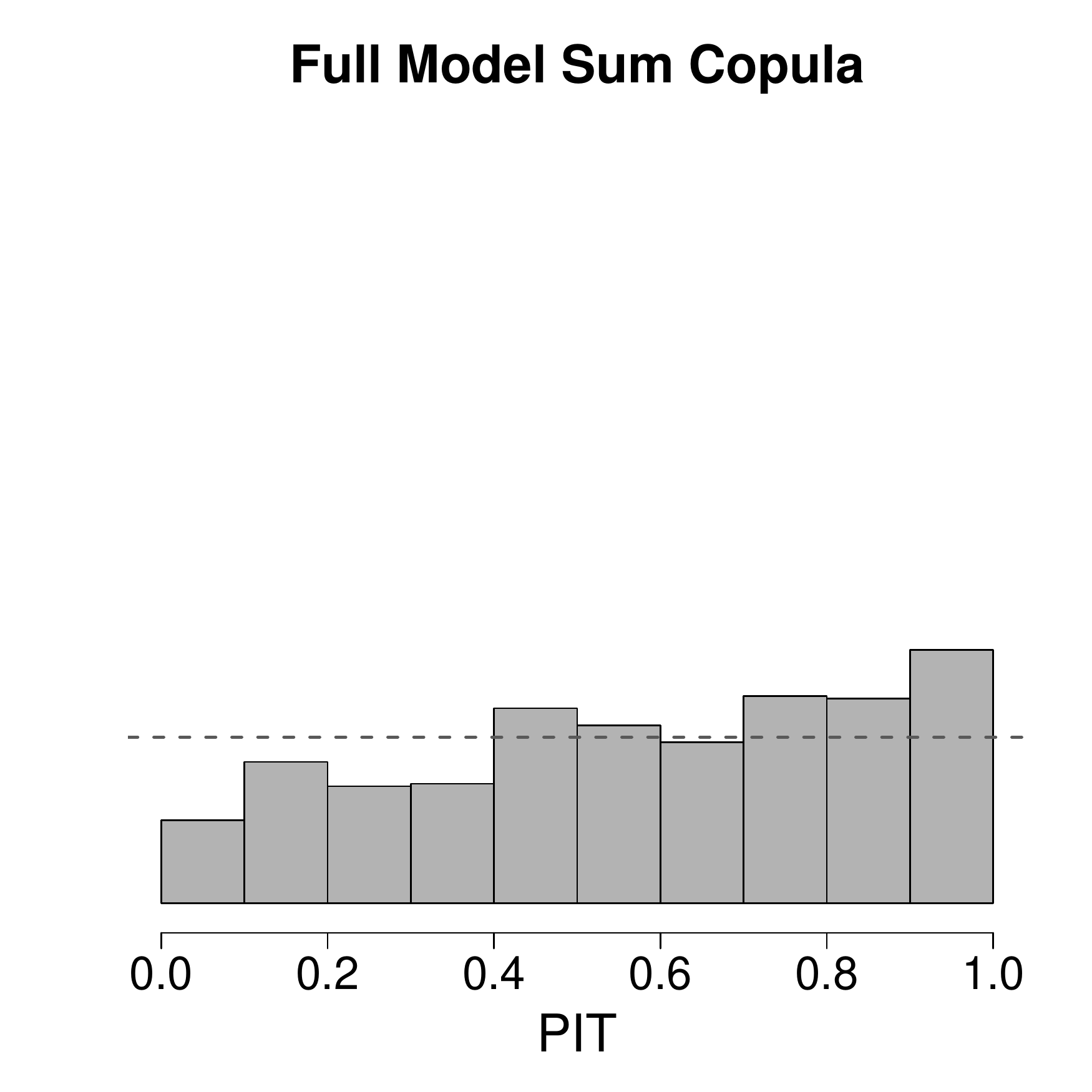}
\includegraphics[width=0.25\textwidth]{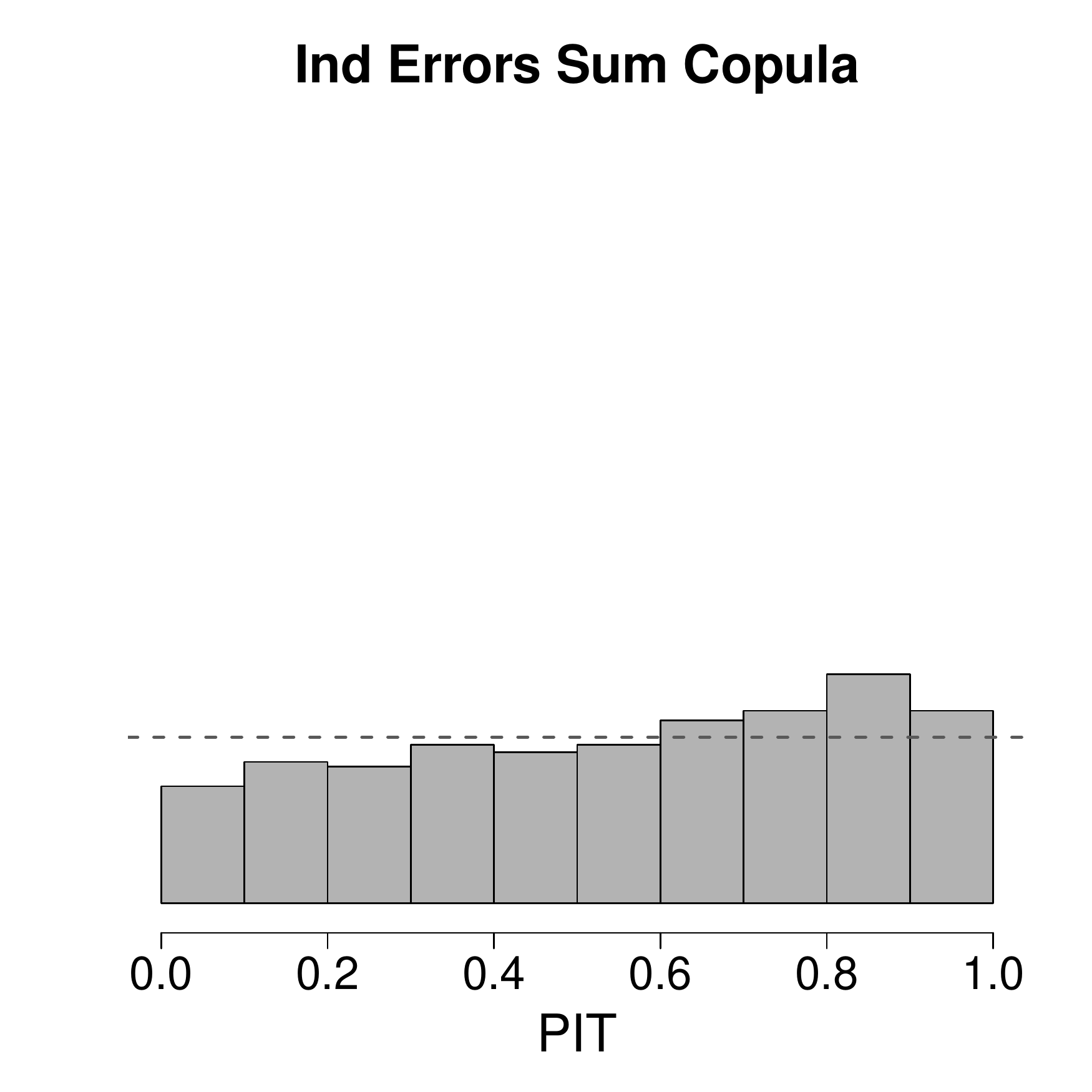}
\includegraphics[width=0.25\textwidth]{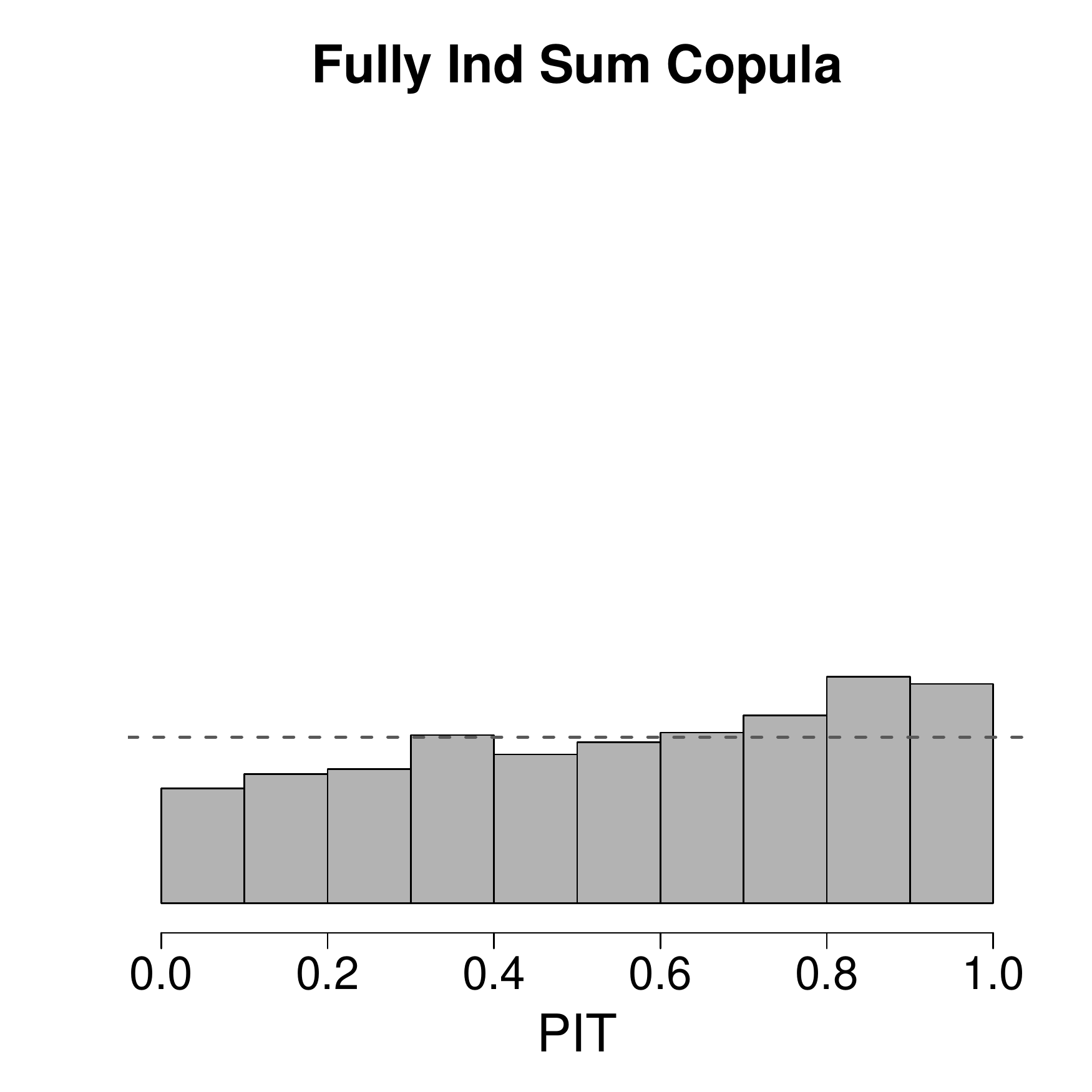}
\caption{PIT for the sum by model type (full, independent errors and fully independent respectively going from left to rigth) and under either the univariate (top row) or copula (bottom row) approach.}\label{fig:hist_sum_wind}
\end{figure}

Figure~\ref{fig:hist_sum_wind} shows the PIT histograms for the total wind forecast under the various model combinations.  In general, even the best cases we seem some evidence of a slight downward bias.  However it is clear that the independence errors approaches are substantially under-dispersive, a behavior that would be expected when the joint distribution is uncorrelated.

\begin{figure}[!hbpt]
\centering
\includegraphics[width=0.25\textwidth]{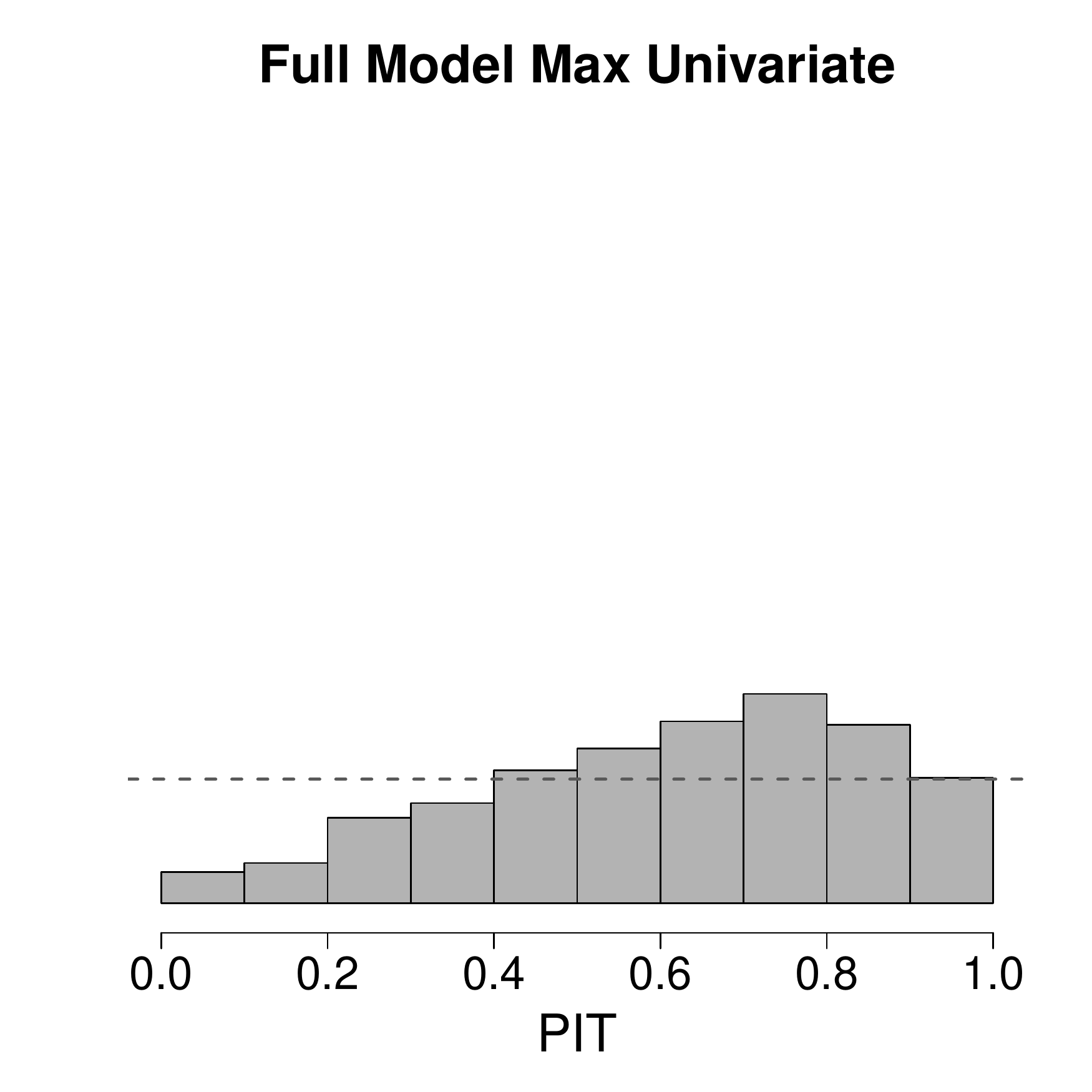}
\includegraphics[width=0.25\textwidth]{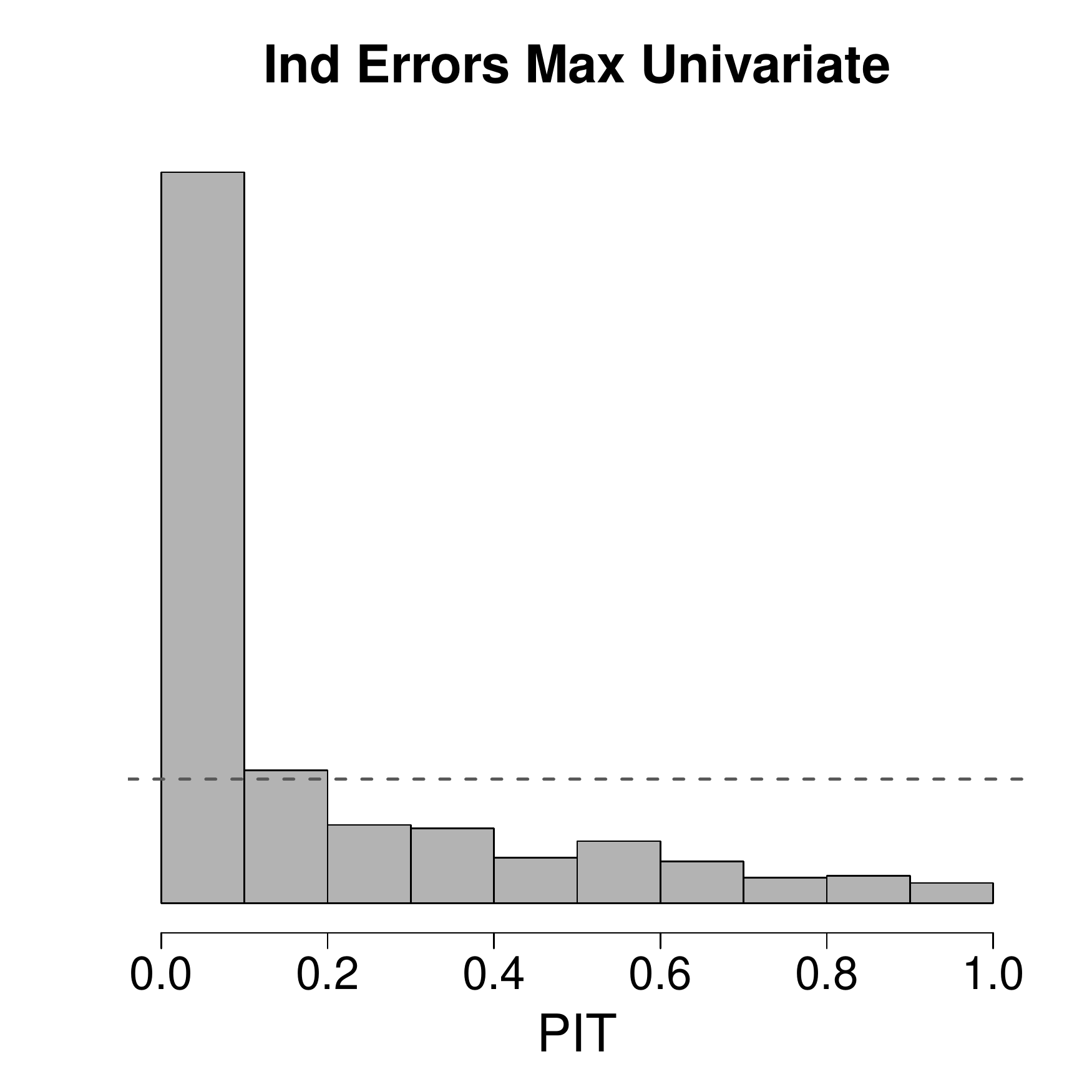}
\includegraphics[width=0.25\textwidth]{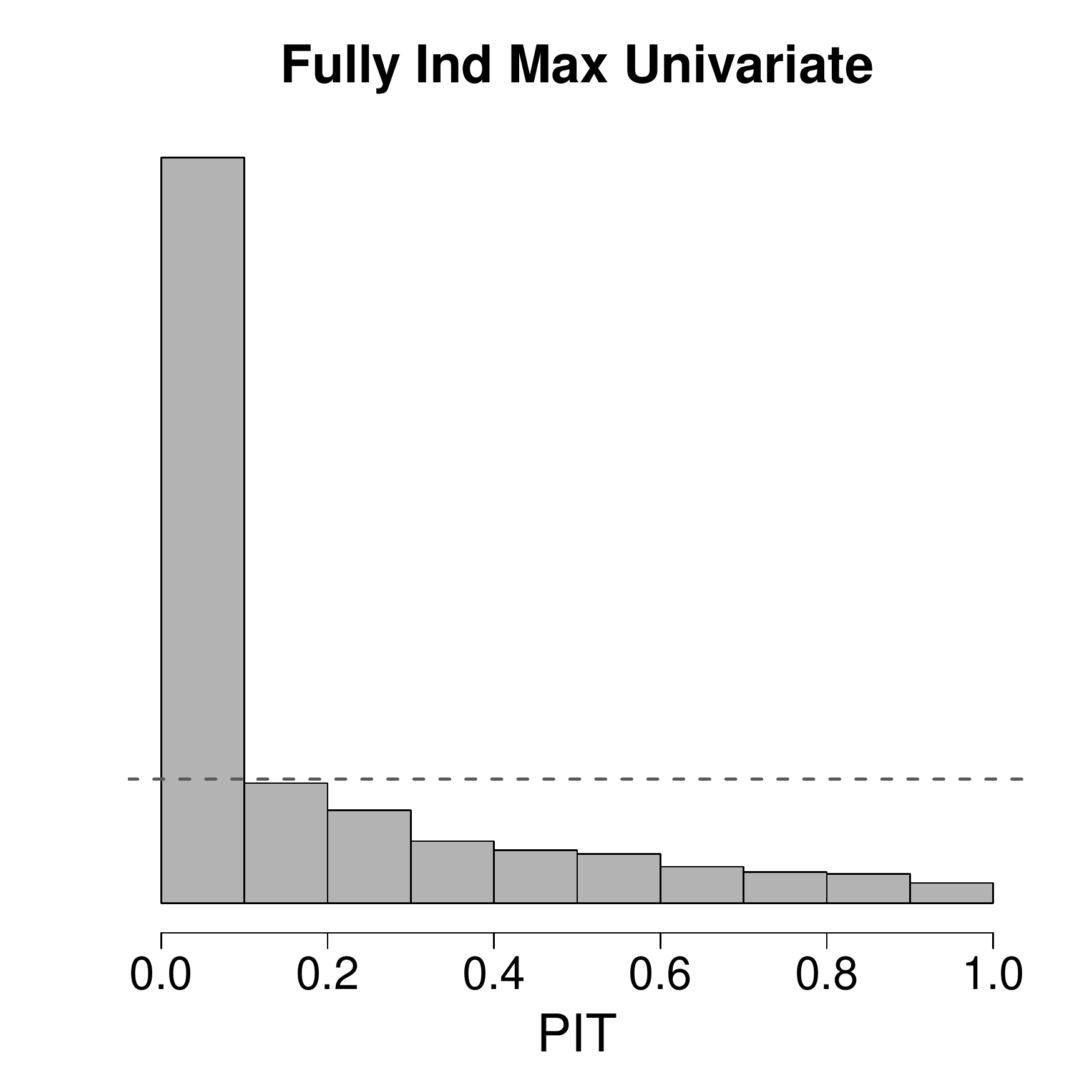}\\
\includegraphics[width=0.25\textwidth]{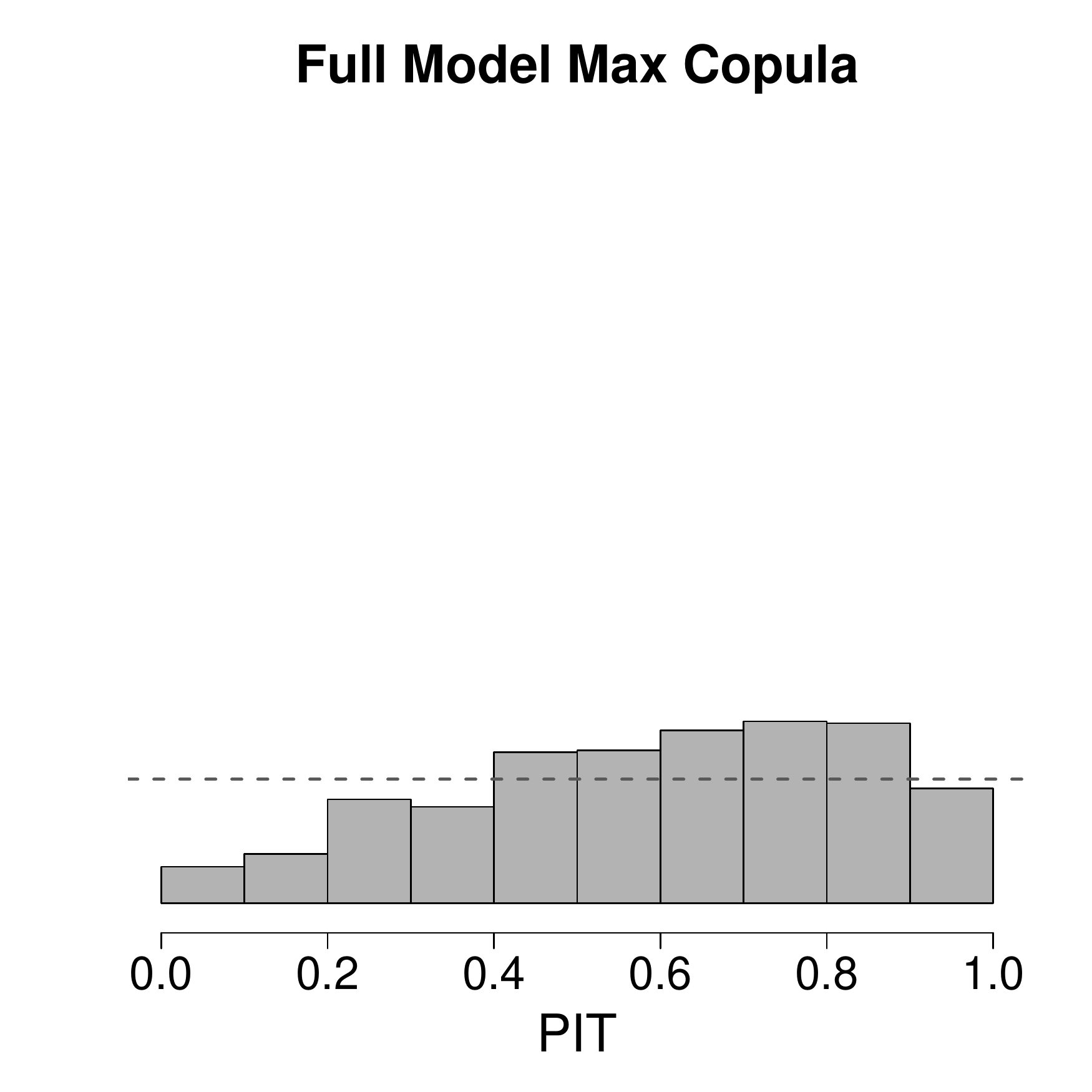}
\includegraphics[width=0.25\textwidth]{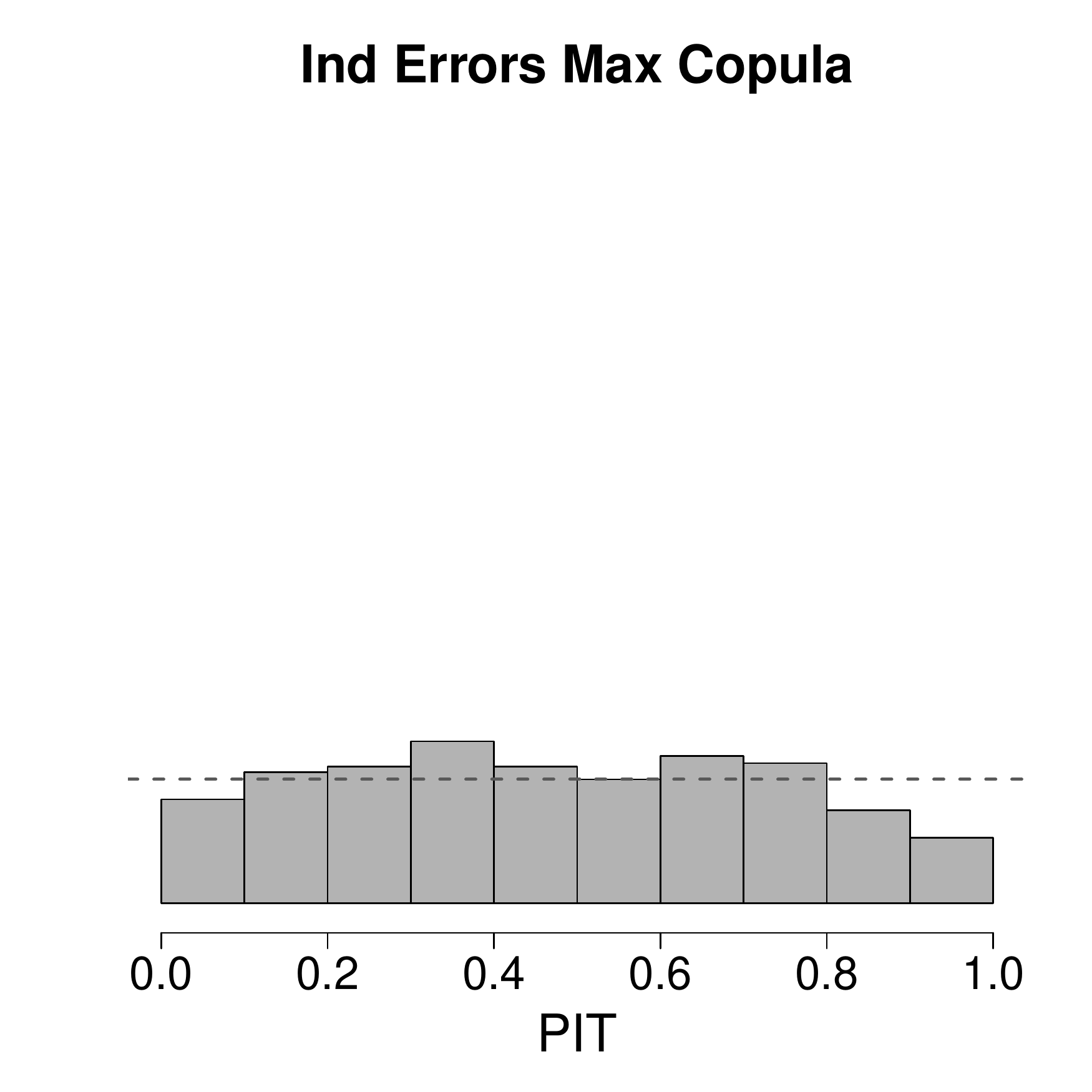}
\includegraphics[width=0.25\textwidth]{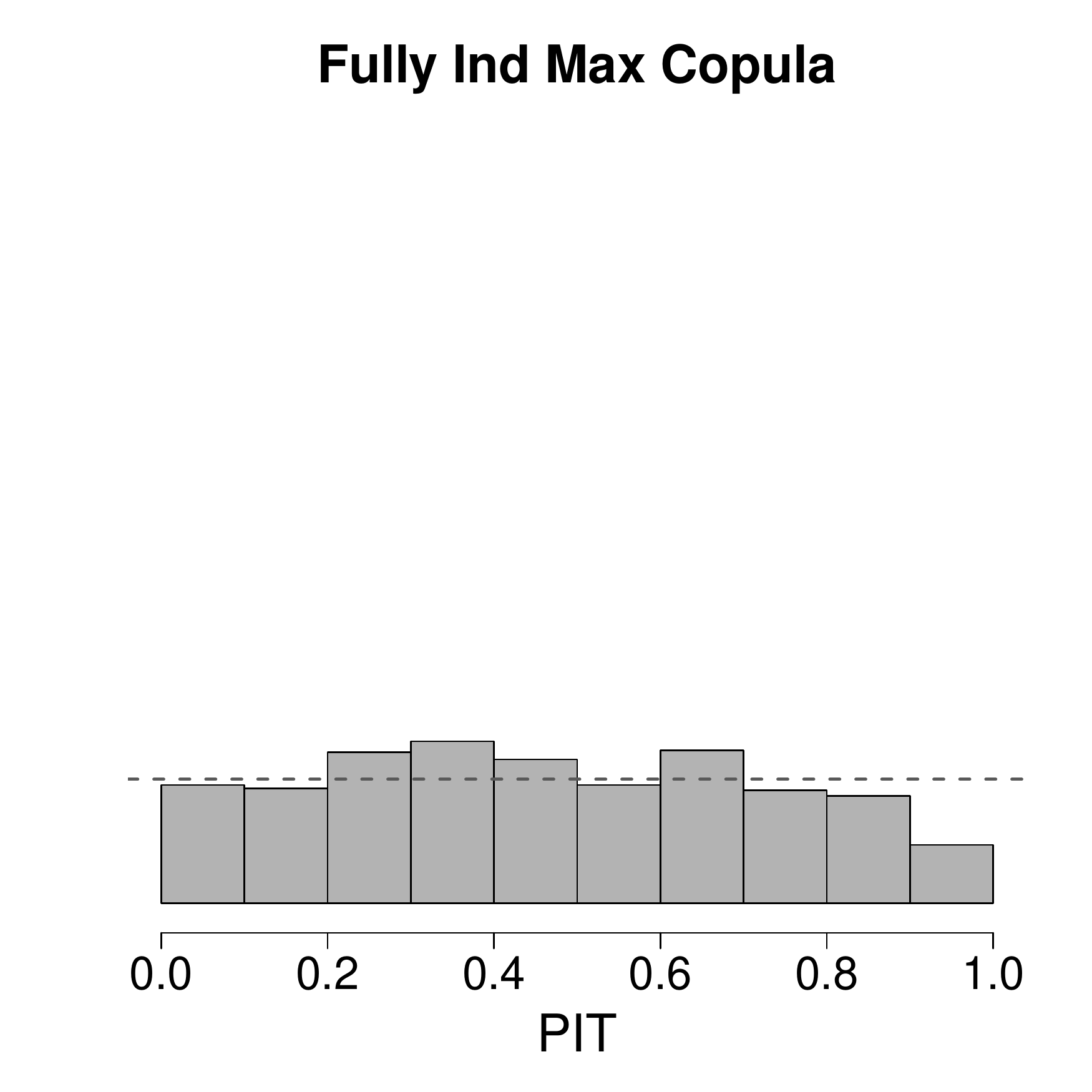}
\caption{PIT for the max by model type (full, independent errors and fully independent respectively going from left to rigth) and under either the univariate (top row) or copula (bottom row) approach.}\label{fig:hist_max_wind}
\end{figure}

Figure~\ref{fig:hist_max_wind} shows the PIT histograms for the maximum wind-speed over 72 hours. In this case, we again see that the independent errors distribution are biased upwards, the actual maximum tends to fall in the lowest quantiles of the predictive distribution.  The copula post-processing of the non-full models appear to have the highest degree of calibration, though potentially too much dispersion at the upper end of the prediction interval.

In general, the results in Tables~\ref{tab:mult_sum}~and~\ref{tab:mult_max} alongside the calibration results in Figures~\ref{fig:hist_sum_wind}~and~\ref{fig:hist_max_wind} show that our approach to joint distributional modeling works well.  The results are conclusive on the usefulness of incorporating hour-to-hour dependence in the regression coefficients.  Further, it is clear that a joint sampling distribution is an important component in achieving calibration.

\section{Discussion\label{sec:discussion}}
We have built a hierarchical Bayesian model for issuing joint distributional forecasts of wind power production in Germany.  This system uses the output of a numerical weather predicition model to derive a predictive feature set.

Our results are clear in the necessity for a joint predictive distribution.  However, we have shown that copula post-processing of marginal forecasts can be a competitive alternative to building a direct full model.

Since neighboring hours are likely to translate features into production estimates in a similar manner, we introduced dependence between regression coefficients via a G-Wishart prior distribution.  Our results show that this approach yields estimates which are sharp and calibrated, both on the univariate and multivariate scales and outperform approaches that use an independent prior on the regression parameters.

There are number of technical manners by which the model could be embellished.  For instance, at the moment the NWP output in a given hour is used to form the features for that particular hour.  There is reason to believe that sharing this information across hours could be beneficial.  More flexible representations than a linear function relating NWP output and wind power production could also be entertained, e.g. smoothing splines which still enable the methodology used here on an expanded feature set.

\section*{Acknowledgment}

This work was performed within Big Insight -- Centre for Research-based Innovation with support from The Research Council of Norway through grant nr. 237718. We thank Stefan Erath from Norsk Hydro for sharing his expertise and data.


\begin{thebibliography}{10}

\bibitem{Bremnes2004}
J~B Bremnes.
\newblock Probabilistic wind power forecasts using local quantile regression.
\newblock {\em Wind Energy}, 7:47--54, 2004.

\bibitem{Dawid1984}
A~P Dawid.
\newblock Statistical theory: The prequential approach (with discussion and
  rejoinder).
\newblock {\em Journal of the Royal Statistical Society Ser.~A}, 147:278--292,
  1984.

\bibitem{DobraLenkoski2011}
A~Dobra and A~Lenkoski.
\newblock Copula {G}aussian graphical models and their application to modeling
  functional disability data.
\newblock {\em Annals of Applied Statistics}, 5:969--993, 2011.

\bibitem{Dowell&2016}
J~Dowell and P~Pinson.
\newblock Very-short-term probabilistic wind power forecasts by sparse vector
  autoregression.
\newblock {\em IEEE Transactions on Smart Grid}, 7(2):763--770, 2016.

\bibitem{elberg&2015}
C~Elberg and S~Hagspiel.
\newblock Spatial dependencies of wind power and interrelations with spot price
  dynamics.
\newblock {\em European Journal of Operational Research}, 241(1):260--272,
  2015.

\bibitem{Engeland&2017}
K~Engeland, M~Borga, J-D~ Creutin, B~
  Fran{\c{c}}ois, M-H~Ramos, and J-P~Vidal.
\newblock Space-time variability of climate variables and intermittent
  renewable electricity production--a review.
\newblock {\em Renewable and Sustainable Energy Reviews}, 79:600--617, 2017.

\bibitem{giebel2011state}
G~Giebel, R~Brownsword, G~Kariniotakis, M~Denhard, and
  C~Draxl.
\newblock The state-of-the-art in short-term prediction of wind power: A
  literature overview, 2nd edition.
\newblock Technical report, ANEMOS.plus, 2011.
\newblock Available at
  \url{http://www.risoe.dtu.dk/rispubl/NEI/NEI-DK-5521.pdf}.

\bibitem{GIEBEL201759}
G~Giebel and G~Kariniotakis.
\newblock 3 - wind power forecasting—a review of the state of the art.
\newblock In George Kariniotakis, editor, {\em Renewable Energy Forecasting},
  pages 59 -- 109. Woodhead Publishing, 2017.

\bibitem{Gneiting2011}
T~Gneiting.
\newblock Making and evaluating point forecasts.
\newblock {\em Journal of the American Statistical Association}, 106:746--762,
  2011.

\bibitem{GneitingBalabdaouiRaftery2007}
T~Gneiting, F~Balabdaoui, and A~E Raftery.
\newblock Probabilistic forecasts, calibration and sharpness.
\newblock {\em Journal of the Royal Statistical Society Ser.~B}, 69:243--268,
  2007.

\bibitem{Gneiting&2006}
T~Gneiting, K~Larson, K~Westrick, M~G Genton, and E~Aldrich.
\newblock Calibrated probabilistic forecasting at the {S}tateline wind energy
  center: {T}he regime-switching space-time method.
\newblock {\em Journal of the American Statistical Association},
  101(475):968--979, 2006.

\bibitem{GneitingRaftery2007}
T~Gneiting and A~E Raftery.
\newblock Strictly proper scoring rules, prediction, and estimation.
\newblock {\em Journal of the American Statistical Association}, 102:359--378,
  2007.

\bibitem{Hagspiel&2012}
S~Hagspiel, A~Papaemannouil, M~Schmid, and G~Andersson.
\newblock Copula-based modeling of stochastic wind power in {E}urope and
  implications for the {S}wiss power grid.
\newblock {\em Applied Energy}, 96:33--44, 2012.

\bibitem{HeringGenton2010}
A~S Hering and M~G Genton.
\newblock Powering up with space-time wind forecasting.
\newblock {\em Journal of the American Statistical Association},
  105(489):92--104, 2010.

\bibitem{Hong&2014}
T~Hong, P~Pinson, and S~Fan.
\newblock Global energy forecasting competition 2012.
\newblock {\em International Journal of Forecasting}, 30(2):357--363, 2014.

\bibitem{Hong&2016}
  T~Hong, P~Pinson, S~Fan, H~Zareipour, A~Troccoli, and
  R~J~Hyndman.
\newblock Probabilistic energy forecasting: Global energy forecasting
  competition 2014 and beyond.
\newblock {\em International Journal of Forecasting}, 32(3):896--913, 2016.

\bibitem{JeonTaylor2012}
J~Jeon and J~W Taylor.
\newblock Using conditional kernel density estimation for wind power density
  forecasting.
\newblock {\em Journal of the American Statistical Association},
  107(497):66--79, 2012.

\bibitem{Jordan&2018}
A~Jordan, F~Krueger, and S~Lerch.
\newblock Evaluating probabilistic forecasts with scoringrules.
\newblock {\em Journal of Statistical Software}, 2018.
\newblock forthcoming.

\bibitem{Kruger&2016}
F~Kr\"uger, S~Lerch, T~L Thorarinsdottir, and T~Gneiting.
\newblock Probabilistic forecasting and comparative model assessment based on
  {M}arkov chain {M}onte {C}arlo output.
\newblock arXiv:1608.06802.

\bibitem{LangeFocken2006}
M~Lange and U~Focken.
\newblock {\em Physical approach to short-term wind power prediction}.
\newblock Springer: Heidelberg, 2006.

\bibitem{Lenkoski2013}
A~Lenkoski.
\newblock A direct sampler for {G}-{Wishart} variates.
\newblock {\em Stat}, 9:119--128, 2013.

\bibitem{LeutbecherPalmer2008}
M~Leutbecher and T~N Palmer.
\newblock Ensemble forecasting.
\newblock {\em Journal of Computational Physics}, 227:3515--3539, 2008.

\bibitem{Louie2014}
H~Louie.
\newblock Evaluation of bivariate {A}rchimedean and elliptical copulas to model
  wind power dependency structures.
\newblock {\em Wind Energy}, 17(2):225--240, 2014.

\bibitem{Messner&2013}
J~W Messner, A~Zeileis, J~Broecker, and G~J Mayr.
\newblock Probabilistic wind power forecasts with an inverse power curve
  transformation and censored regression.
\newblock {\em Wind Energy}, 2013.
\newblock DOI: 10.1002/we.1666.

\bibitem{Moeller&2013}
A~M{\"o}ller, A~Lenkoski, and T~L Thorarinsdottir.
\newblock Multivariate probabilistic forecasting using {B}ayesian model
  averaging and copulas.
\newblock {\em Q J Roy Meteor Soc}, 139(673):982--991, 2013.

\bibitem{Molteni&1996}
A~Molteni, R~Buizza, T~N Palmer, and T~Petroliagis.
\newblock The new {ECMWF} ensemble prediction system: Methodology and
  validation.
\newblock {\em Quarterly journal of the royal meteorological society},
  122:73--119, 1996.

\bibitem{nagy&2016}
  G~I~Nagy, G~Barta, S~Kazi, G~Borb{\'e}ly, and
  G~Simon.
\newblock Gefcom2014: Probabilistic solar and wind power forecasting using a
  generalized additive tree ensemble approach.
\newblock {\em International Journal of Forecasting}, 32(3):1087--1093, 2016.

\bibitem{PapaefthymiouKurowicka2009}
G~Papaefthymiou and D~Kurowicka.
\newblock Using copulas for modeling stochastic dependence in power system
  uncertainty analysis.
\newblock {\em IEEE Transactions on Power Systems}, 24(1):40--49, 2009.

\bibitem{Pinson2012}
P~Pinson.
\newblock Very-short-term probabilistic forecasing of wind power with
  generalized logit-normal distributions.
\newblock {\em Journal of the Royal Statistical Society Series C},
  61(4):555--576, 2012.

\bibitem{Pinson2013}
P~Pinson.
\newblock Wind energy: {F}orecasting challenges for its operational management.
\newblock {\em Statistical Science}, 28(4):564--585, 2013.

\bibitem{PinsonKariniotakis2010}
P~Pinson and G~N Kariniotakis.
\newblock Conditional prediction intervals of wind power generation.
\newblock {\em IEEE Transactions on Power Systems}, 25:1845--1856, 2010.

\bibitem{PinsonMadsen2009}
P~Pinson and H~Madsen.
\newblock Ensemble-based probabilistic forecasting at {H}orns {R}ev.
\newblock {\em Wind Energy}, 12(2):137--155, 2009.

\bibitem{Pinson&2009}
P~Pinson, H~Madsen, H~A Nielsen, G~Papaefthymiou, and B~Klöckl.
\newblock From probabilistic forecasts to statistical scenarios of short-term
  wind power production.
\newblock {\em Wind Energy}, 12(1):51--62, 2009.

\bibitem{Roverato2002}
A~Roverate.
\newblock Hyper inverse {W}ishart distribution for non-decomposable graphs and
  its applications to {B}ayesian inference for {G}aussian graphical models.
\newblock {\em Scandinavian Journal of Statistics}, 29:391--411, 2002.

\bibitem{SiebertKariniotakis2006}
N~Siebert and G~Kariniotakis.
\newblock Reference wind farm selection for regional wind power prediction
  models.
\newblock In {\em Proceedings of the European Wind Energy Conference, EWEC
  2006}, 2006.

\bibitem{Taylor&2009}
J~W Taylor, P~E {McSharry}, and R~Buizza.
\newblock Wind power density forecasting using ensemble predictions and time
  series models.
\newblock {\em IEEE Transactions on Energy Conversion}, 24(3):775--782, 2009.

\bibitem{Thorarinsdottir&2016}
T~L Thorarinsdottir, M~Scheuerer, and C~Heinz.
\newblock Assessing the calibration of high-dimensional ensemble forecasts
  using rank histograms.
\newblock {\em Journal of Computational and Graphical Statistics},
  25(1):105--122, 2013.

\bibitem{zhang2016}
Y~Zhang and J~Wang.
\newblock K-nearest neighbors and a kernel density estimator for gefcom2014
  probabilistic wind power forecasting.
\newblock {\em International Journal of Forecasting}, 32(3):1074--1080, 2016.

\end{thebibliography}
\end{document}